\documentclass[12pt,preprint]{aastex}

\usepackage{graphicx,natbib}

\begin{document}

\newcommand{\ndla}{71}
\newcommand{\kms}{km~s$^{-1}$ }
\newcommand{\cm}[1]{\, {\rm cm^{#1}}}
\newcommand{\mkms}{{\rm \; km\;s^{-1}}}
\newcommand{\delv}{$\Delta v_{90}$}
\newcommand{\hdelv}{$\Delta v_{90}^{\rm high}$}
\newcommand{\wsi}{$W_{\rm 1526}$}
\newcommand{\wciv}{$W_{\rm CIV}$}
\newcommand{\mwciv}{W_{\rm CIV}}
\newcommand{\ohi}{$\Omega_g$}
\newcommand{\lya}{Ly$\alpha$}
\newcommand{\nv}{N\,V}
\newcommand{\ovi}{O\,VI}
\newcommand{\N}[1]{{N({\rm #1})}}
\newcommand{\sci}[1]{{\rm \; \times \; 10^{#1}}}
\newcommand{\mnhi}{N_{\rm HI}}
\newcommand{\mnciv}{N_{\rm CIV}}
\newcommand{\afeg}{[$\alpha$/Fe]$_g$}
\newcommand{\nhii}{$\N{H^+}$}
\newcommand{\nhi}{$N_{\rm HI}$}
\newcommand{\nhtwo}{$N_{\rm H_2}$}
\def\fnhi{$f(\mnhi)$}
\def\mfnhi{f(\mnhi)}
\def\ltk{\left [ \,}
\def\ltp{\left ( \,}
\def\ltb{\left \{ \,}
\def\rtk{\, \right  ] }
\def\rtp{\, \right  ) }
\def\rtb{\, \right \} }
\def\lnhi{$\log N_{HI}$}
\def\omt{$\Omega_m^{Total}$}
\def\momt{\Omega_m^{Total}}

\title{The Rapidly Flaring Afterglow of the Very Bright and Energetic GRB\,070125}

\author{Adria C. Updike\altaffilmark{1}, Josh B. Haislip\altaffilmark{2}, 
Melissa C. Nysewander\altaffilmark{3}, Andrew S. Fruchter\altaffilmark{3},
D. Alexander Kann\altaffilmark{4}, Sylvio Klose\altaffilmark{4},
Peter A. Milne\altaffilmark{5}, 
G. Grant Williams\altaffilmark{6}, Weikang Zheng\altaffilmark{7},
Carl W. Hergenrother\altaffilmark{8},
Jason X. Prochaska\altaffilmark{9},  Jules P. Halpern\altaffilmark{10},
Nestor Mirabal\altaffilmark{10}, John R. Thorstensen\altaffilmark{11},
Alexander J. van der Horst\altaffilmark{12}, 
Rhaana L. C. Starling\altaffilmark{13},
Judith L. Racusin\altaffilmark{14},  David N. Burrows\altaffilmark{14},
N. P. M. Kuin\altaffilmark{15}, Peter W. A. Roming\altaffilmark{14},
Eric Bellm\altaffilmark{16}, Kevin Hurley\altaffilmark{16},
Weidong Li\altaffilmark{17}, Alexei V. Filippenko\altaffilmark{17},
Cullen Blake\altaffilmark{18},  Dan Starr\altaffilmark{17}, 
Emilio E. Falco\altaffilmark{19}, Warren R. Brown\altaffilmark{18}, 
Xinyu Dai\altaffilmark{20}, 
 Jinsong Deng\altaffilmark{7},
 Liping Xin\altaffilmark{7},
Yulei Qiu\altaffilmark{7}, Jianyan Wei\altaffilmark{7},
Yuji Urata\altaffilmark{21,22},
Domenico Nanni\altaffilmark{23,24}, 
 Elisabetta Maiorano\altaffilmark{25}, Eliana Palazzi\altaffilmark{25},
 Giuseppe Greco\altaffilmark{26}, Corrado Bartolini\altaffilmark{26},
Adriano Guarnieri\altaffilmark{26}, Adalberto Piccioni\altaffilmark{26},
Graziella Pizzichini\altaffilmark{25}, Federica Terra\altaffilmark{23},
  Kuntal Misra\altaffilmark{27,28},
B. C. Bhatt\altaffilmark{29}, G. C. Anupama\altaffilmark{30}
X. Fan\altaffilmark{5},
L. Jiang\altaffilmark{5}, 
Ralph A. M. J. Wijers\altaffilmark{31}, 
Daniel E. Reichart\altaffilmark{2},
Hala A. Eid\altaffilmark{1}, Ginger Bryngelson\altaffilmark{1},
Jason Puls\altaffilmark{1}, R. C. Goldthwaite\altaffilmark{1},
and Dieter H. Hartmann\altaffilmark{1}
}

\altaffiltext{1}{Department of Physics and Astronomy, Clemson University 
118 Kinard Laboratory, Clemson, SC 29634.}
\altaffiltext{2}{Department of Physics and Astronomy, University 
of North Carolina at Chapel Hill, Campus Box 3255, Chapel Hill, NC 27599.}
\altaffiltext{3}{Space Telescope Science Institute, 3700 San
Martin Dr., Baltimore, MD 21218.}
\altaffiltext{4}{Th\"uringer Landessternwarte Tautenburg,  
Sternwarte 5, D--07778 Tautenburg, Germany.}
\altaffiltext{5}{Steward Observatory, University of Arizona, Tucson
AZ 85721-0065.}
\altaffiltext{6}{MMT Observatory, University of Arizona, 
Tucson AZ 85721-0065.}
\altaffiltext{7}{National Astronomical Observatories, 20A Datun Rd., 
Chaoyang Dist., Beijing 100012, China.}
\altaffiltext{8}{Lunar and Planetary Laboratory, University
of Arizona, Tucson AZ 85721-0092.}
\altaffiltext{9}{UCO/Lick Observatory, University of California,
Santa Cruz, CA 95064.}
\altaffiltext{10}{Columbia Astrophysics Laboratory, Columbia University,
550 West 120th Street, New York, NY 10027.}
\altaffiltext{11}{Dept. of Physics and Astronomy, Dartmouth College
6127 Wilder Laboratory
Hanover, NH 03755-3528.}
\altaffiltext{12}{NASA Postdoctoral Program Fellow, NSSTC, 320 Sparkman Drive, Huntsville, AL 35805.}
\altaffiltext{13}{Department of Physics and Astronomy, University of Leicester, University Road, Leicester, LE1 7RH, UK.}
\altaffiltext{14}{Penn State University, Department of Astronomy
\& Astrophysics, 525 Davey Lab, University Park, PA 16802.}
\altaffiltext{15}{Mullard Space Science Laboratory/UCL, Holmbury St. Mary, Dorking, Surrey, HR5, 6NT, UK.}
\altaffiltext{16}{University of California, Space Sciences Laboratory, 7 Gauss Way, Berkeley, CA 94720-7450.}
\altaffiltext{17}{Department of Astronomy, University of California,
Berkeley, CA 94720-3411.}
\altaffiltext{18}{Harvard-Smithsonian Center for Astrophysics,
60 Garden Street, MS-20, Cambridge, MA 02138.}
\altaffiltext{19}{Smithsonian Institution, Whipple Observatory,
670 Mt. Hopkins Road, P.O. Box 6369, Amado, AZ 85645.}
\altaffiltext{20}{Ohio State University Department of Physics,
191 W. Woodruff Ave, Columbus, OH 43210-1117.}
\altaffiltext{21}{Dept. of Physics, Saitama University,
Shimookubo, Urawa 338-8570 Japan.}
\altaffiltext{22}{Academia Sinica Institute of Astronomy and Astrophysics,
Taipei 106, Taiwan, Republic of China.}
\altaffiltext{23}{Second University of Roma "Tor Vergata," Italy.}
\altaffiltext{24}{INAF/OAR.}
\altaffiltext{25}{INAF/IASF Bologna, via Gobetti 101, 40129 Bologna, Italy.}
\altaffiltext{26}{Dipartimento di Astronomia, Universite di Bologna, 
via Ranzani 1, 40127  Bologna, Italy.}
\altaffiltext{27}{Aryabhatta Research Institute of observational sciencES (ARIES), Manora Peak, Nainital - 263 129 India.}
\altaffiltext{28}{Inter University Center for Astronomy and Astrophysics, Post Bag 4, Ganeshkhind, Pune 411 007 India.}
\altaffiltext{29}{Center for Research and Education in Science and Technology (CREST), Hosakote, Bangalore - 562 114 India.}
\altaffiltext{30}{Indian Institute of Astrophysics, Bangalore - 560 034 India.}
\altaffiltext{31}{Astronomical Institute, University of Amsterdam, Kruislaan 403, 1098 SJ Amsterdam, The Netherlands.}

\begin{abstract}

We report on multi-wavelength observations, ranging 
from X-ray to radio wave bands, of the IPN-localized 
gamma-ray burst GRB 070125.  Spectroscopic observations reveal the presence 
of absorption lines due to O~I, Si~II, and C~IV, implying a likely 
redshift of $z$ = 1.547.   The well-sampled light curves,
in particular from 0.5 to 4 days after the burst,
suggest a jet break at 3.7
days, corresponding to a jet opening angle of $\sim$7.0$^\circ$, and
implying an intrinsic GRB energy in the 1--10,000 keV band of around 
$E_{\gamma} = (6.3-6.9) \times 10^{51}$ erg (based on the 
fluences measured by the gamma-ray detectors of
the IPN network). GRB\,070125 is among the brightest 
afterglows observed to date.  The spectral energy distribution implies 
a host extinction of $A_{V} < 0.9$ mag.   
Two rebrightening episodes are observed, one with  
excellent time coverage, showing an increase in flux of 56\% in  
$\sim$8000 seconds.  The evolution of the afterglow 
light curve is achromatic at all times.  
Late-time observations of the afterglow do not show evidence 
for emission from an underlying host galaxy or supernova.  Any host 
galaxy would be subluminous, consistent with current GRB  
host-galaxy samples.  Evidence for strong Mg~II absorption features 
is not found, which is  
perhaps surprising in view of the relatively  
high redshift of this burst and the 
high likelihood for such features  
along GRB-selected lines of sight. 
 
\end{abstract} 
 
 
\section{Introduction}   

Since the detection of the first gamma-ray burst (GRB) in 1967 with detectors  
aboard the Vela satellites \citep{1973ApJ...182L..85K}, our understanding of  
the nature of this still somewhat mysterious phenomenon went through several  
major advances. In particular, the establishment of their cosmological  
distances via ground-based follow-up spectroscopy of their afterglow emission,  
first accomplished for GRB 970508 \citep{1997Natur.387..878M} at a redshift  
of $z$ = 0.835, led to a spectacular world-wide effort to accumulate  
prompt and afterglow observations which have yielded many surprises  
and breakthrough discoveries. One decade after the first afterglow was  
discovered with BeppoSAX \citep{1997Natur.387..783C, 1997Natur.386..686V},  
the accumulated sample of about 500 GRBs\footnote{See the compilation  
in http://www.mpe.mpg.de/\symbol{126}jcg/grbgen.html} exhibits   
X-ray, optical, and radio afterglows to varying degrees.  
Their observed redshift distribution \citep{2006A&A...447..897J}  
is very broad, with more than half of the GRBs at distances beyond  
the peak of the cosmic star formation rate at $z$ $\sim$ $1-2$, and  
with GRB 050904 at $z$ = 6.29 currently being the most  
distant burst \citep{2006Natur.440..181H, 2006Natur.440..184K}. The  
association of long-duration GRBs with Type Ib/c supernovae (see 
\citealt{1997ARA&A..35..309F} for a review of supernova classification), 
established  
in a few cases via direct spectroscopy and in a larger sample  
via a late extra emission component revealed in broad-band photometric  
observations (e.g., \citealt{1998Natur.395..670G,  
2003Natur.423..847H, 2003ApJ...591L..17S, 2004ApJ...609..952Z,  
2004ApJ...609L...5M, 2006Natur.442.1011P, 2006ApJ...643L..99M,  
2006ARA&A..44..507W}) has thus opened promising observational  
windows into star formation in the universe  
(e.g., \citealt{2006ApJ...642..382B}), cosmic chemical  
evolution via absorption-line  
spectroscopy of intervening clouds and host galaxies (e.g.,  
\citealt{2006NJPh....8..195S, 2006ApJ...642..979B}),   
and may eventually provide valuable constraints on the cosmic  
re-ionization history in  
the crucial $z>6$ epoch (e.g., \citealt{2006ARA&A..44..415F,  
2006PASJ...58..485T, 
2007arXiv0710.1303G, 2007arXiv0710.1018M}).  
Every new GRB provides yet another opportunity to investigate these  
topics, or the  
GRB environment via features in the afterglow power-law decay, or to add  
key data to enable  
a better understanding of their statistical properties and morphological  
classification. 
 
Here we report on follow-up observations of GRB\,070125,  
which was discovered by detectors aboard  
the IPN members Mars Odyssey, Suzaku, INTEGRAL,  
Konus-Wind, and RHESSI at $T_0$ = 07:20:42 (UT dates and times 
are used throughout this paper) on January 25, 2007 
\citep{2007GCN..6024....1H}. 
The BAT detector aboard \emph{Swift} \citep{2004ApJ...611.1005G}  
recorded the burst during a  
slew, and therefore did not trigger.  
The BAT position is consistent with  
the IPN localization.  
The position was  monitored with the  
XRT from 0.54 to 18.5 days after $T_0$.  We  
report these observations and their implications  
(see also \citealt{2007GCNR...28....3R}), as  
well as the results obtained from ground-based  
follow-up observations with a large group of  
small, mid-sized, and large-aperture telescopes.  
As observed by RHESSI, GRB 070125 had a  
duration of $T_{90}$=63.0 $\pm$ 1.7 seconds \citep{2007arXiv0710.4590B},  
and thus clearly belongs in the class of  
long-(soft) GRBs (e.g., \citealt{1993ApJ...413L.101K}).  
Details of the prompt emission can be found in  
\cite{2007arXiv0710.4590B}.  From a joint fit to the 
RHESSI and Konus data, \cite{2007arXiv0710.4590B} 
derived isotropic energies in the 1 keV $-$ 10 MeV band of  
$(9.44^{+0.40}_{-0.41}) \times 10^{53}$ erg (Konus)  
and $(8.27 \pm 0.39) \times 10^{53}$ erg (RHESSI). 
 
Starting with our early response using the 0.9~m  
SARA telescope and the 2.3~m Bok telescope  
on Kitt Peak, we compile observations from a large set of  
follow-up programs to establish the afterglow light-curve  
properties. We analyze a diverse set of data with  
a uniform analysis method to reduce any 
scatter and establish a well-sampled light curve  
(which is usually not possible with data from a  
single telescope). We obtained Keck-I spectra to  
establish a burst redshift of $z$ = 1.547  
\citep[which agrees with the value derived from  
Gemini-North spectroscopy,][]{2007arXiv0712.2828C}. 
The afterglow initially appeared to have a jet break  
around day 1.5, but the analysis presented here shows  
that at this time multiple rebrightening episodes occur,  
which are in fact better fit with abrupt  
jumps in the flux, and not the usual jet-break.  
However, the available sparse late-time data do 
indicate that a break in fact does take place,  
at $\sim$ 3.7 days after the burst, which agrees with the  
radio light curve at 4.8 GHz presented here, spanning 
a time range from 1.5 to 278 days after the burst.

We discuss how these observations place GRB\,070125  
into the context of other gamma-ray bursts and  
how the late-time observations may  
constrain the properties of the underlying host  
and a potential supernova that is expected to reach  
its light maximum $\sim$ 10(1+$z$) days after gamma-ray  
emergence. 
Emission was detected from this source with  
the LBT at $t$ = 26.8 days  
\citep{2007GCN..6165....1G, 2007arXiv0712.2239D}, but the flux  
is too large to be explained with  
emission from the commonly assumed SN 1998bw-like template  
at the burst redshift of 1.547, and we instead consider the  
LBT detection to be  
the afterglow.  This interpretation is supported by a  
second-epoch LBT observation  ($R > 26.1$ mag)  
which did not result in a detection \citep{2007arXiv0712.2239D}. 
 
The paper is organized as follows. In $\S$2 we  
describe our set of observations of GRB\,070125,  
from early X-ray and UVOT data from \emph{Swift}  
to late-time optical observations with the 8.4m LBT  
and late-time radio observations with the WSRT. 
In $\S$3, we describe the data analysis and results  
derived from the multi-wavelength data, and  
in $\S$4 we discuss the implications of this combined  
data set. In particular, we discuss GRB\,070125  
in the context of the existing data on afterglow  
emission properties, and we examine the implied  
burst energies in light of the standard models  
for long-soft GRBs and their interactions with  
the circumburst medium.

\section{Observations}

GRB\,070125 was detected at 07:20:42 on January 25, 2007, by Mars Odyssey 
(HEND and GRS), Suzaku (WAM) (\citealt{yamaoka1}, Yamaoka 2008 in prep.),  
INTEGRAL (SPI-ACS), KONUS, and RHESSI \citep{2007GCN..6024....1H}. 
The \emph{Swift} satellite detected the burst, but did not trigger because  
\emph{Swift} was slewing. The GRB entered the coded field of view of the 
\emph{Swift} BAT 6 minutes after the trigger time, and ground processing  
revealed a significant source at the intersection of the IPN annuli,  
strongly reducing the error box.  The Burst Alert Telescope  
(BAT) \citep{2005SSRv..120..143B} produced a refined J2000 location of  
$\alpha = 7^{\rm h}\ 51^{\rm m}\ 24^{\rm s}$,  
$\delta = +31^{\circ}\ 08'\ 24''$  based on data taken about  
6 minutes after $T_0$ 
\citep{2007GCNR...28....1R}.  
About 13 hours later, \emph{Swift} began observing  
GRB 070125 as a target of opportunity observation.   
The afterglow was discovered with the Palomar 1.5~m telescope 
\citep{2007GCN..6028....1C} and produced an accurate afterglow location of  
$\alpha = 7^{\rm h}\ 51^{\rm m}\ 17^{\rm s}$,  
$\delta = +31^{\circ}\ 09'\ 04'' $ ($\pm$ 0.5$''$ in each coordinate). 
 
\begin{figure} 
\begin{center} 
\epsscale{0.7}
\plotone{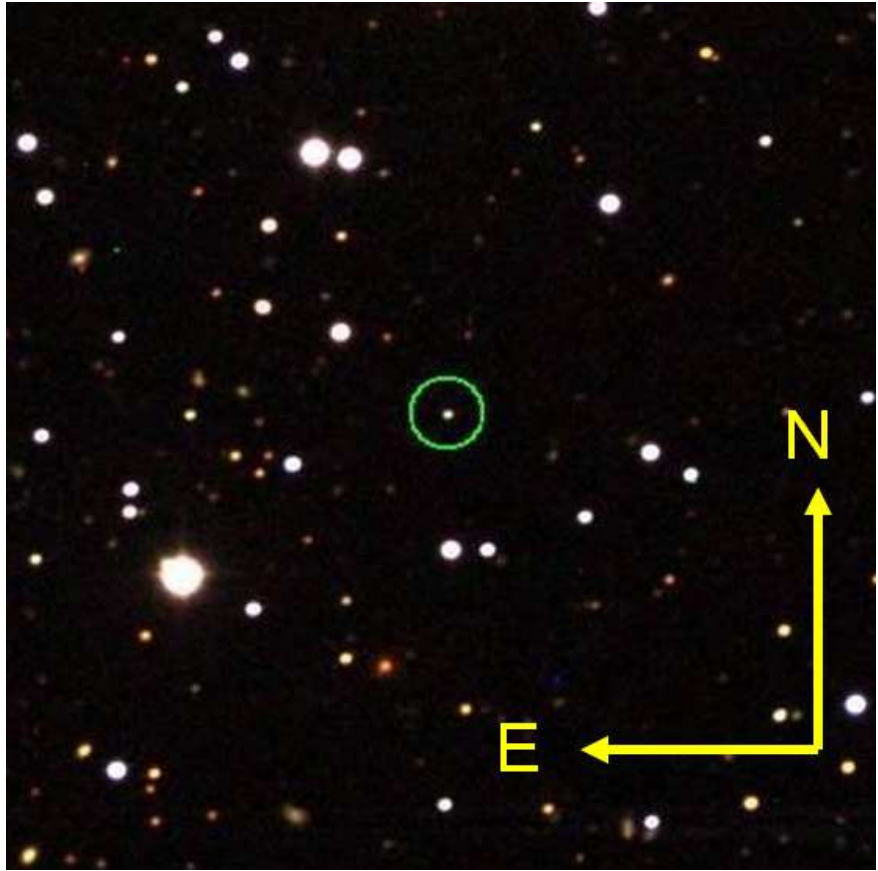} 
\caption{The field of GRB\,070125 as imaged by the Bok telescope on Jan. 26, 
2007, 08:11:08 ($BVR$).  The field is 6.4' in diameter.  The OT is circled. In 
this image, north is up and east is to the left.} 
\label{fig:prompt} 
\end{center} 
\end{figure}

Below, we describe space-based observations  
with instruments on the \emph{Swift} satellite, as well as ground-based  
photometric and spectroscopic observations. Figure~\ref{fig:prompt} shows a  
false-color image of the field derived from $BVR$ observations with the Bok  
telescope (described in greater detail below). The afterglow shown in  
the image  
within the circle has a brightness of $R = 18.7$ mag at one
 day after the burst. The  
field has a relatively low star density 
(Galactic coordinates $l = 189.4^\circ$, $b = 25.6^\circ$) and has 
only a little foreground extinction of $E(B-V) = 0.052$ mag, 
$A_{V} = 0.16$ mag   
\citep{1998ApJ...500..525S}. 
 
\subsection{Space-Based Observations} 
 
The space-based observations reported in this section consist of ultraviolet, 
optical, and X-ray data, carried out by the UVOT and XRT  
instruments on $Swift$  
beginning 12.97 hours after the burst trigger. The afterglow was detected in  
the X-ray band and all six UVOT filters, ranging from $V$ to uvw2 (central 
wavelengths of 546 nm and 193 nm, respectively). 
 
\subsubsection{XRT Observations} 
 
Although GRB\,070125 was detected by the \emph{Swift}-BAT  
in ground analysis, the BAT did 
not trigger due to the occurrence of the burst during a  
pre-planned slew phase, and  
consequently the narrow-field instruments did not  
obtain prompt observations. \emph{Swift}  
began Target of Opportunity observations of GRB\,070125  
about 13 hours after the 
trigger at 20:18:48 January 25, 2007. 
 
The X-ray afterglow was initially detected by the XRT  
\citep{2005SSRv..120..165B} in the  
first few orbits of observation and was followed  
up for 18.5 days post-trigger until the  
afterglow was no longer detected by the XRT. The total  
exposure time of GRB\,070125  
with the XRT was 170 ks. All observations used the 
 Photon Counting (PC) mode due to the low count rate of  
the source. Level-1 data products were downloaded from  
the NASA/GSFC \emph{Swift} Data  
Center (SDC) and processed using XRTDAS software (v2.0.1).  
The {\it xrtpipeline}  
task was used to generate level-2 cleaned event files.  
Only events with PC grades 
0--12 and energies in the range $0.3-10.0$ keV were used in  
subsequent temporal and 
spectral analysis. 
 
The XRT light curve was created by extracting the counts in a circular region 
around the afterglow position with a variable radius designed to optimize 
the signal-to-noise ratio (S/N)  
depending on the count rate. A region with 40 pixels  
radius clear of serendipitous  
background sources was used to estimate the contribution  
of background counts in  
the source extraction region. The number of counts per  
bin was chosen depending on  
the count rate to show sufficient detail with reasonable  
error bars. Background-subtracted  
count rates were also corrected for the  
portion of the PSF excluded by  
the extraction region and any proximity to bad columns  
and hot pixels in the XRT  
CCD. Pile-up is negligible at these flux levels. 
 
We carried out spectral analysis on this data set to obtain a counts-to-flux  
 conversion for the purpose of  creating an XRT light curve in   
standard units.  Spectral analysis was carried out using XSPEC   
(v12.3.1) and the XRT ancillary response file created with the   
standard {\it xrtmkarf} task using the response matrix file   
swxpc0to12s0$\_$20010101v010.rmf from CALDB (release 2007-12-04).  We   
fit the spectrum to a simple absorbed power law with an absorption   
system at $z = 0$ fixed to the Galactic HI column density ($4.8\times10^  
{20}\ {\rm cm}^{-2}$, \citealt{1990ARA&A..28..215D}),  
and allowed the absorption at the   
host-galaxy redshift ($z = 1.547$) to vary.  The resulting fit to the PC   
spectrum gave an intrinsic effective hydrogen column density of   
$N_H=0.19^{+0.20}_{-0.18}\times10^{22}\ {\rm cm}^{-2}$ and  
a photon index of $2.00^  
{+0.15}_{-0.14}$, with a reduced $\chi^2$ of 0.67 (29 degrees of freedom [d.o.f.]).   
The  absorption column is in excess of the Galactic foreground   
value, suggesting the possibility that the host galaxy or the circumburst  
medium contribute somewhat to the observed extinction.  We   
address this issue below with the X-ray/optical/near-IR afterglow  
spectral energy distribution (SED).    
The mean absorbed (un-absorbed) flux is $3.9\times  
10^{-13}\ {\rm erg} \ {\rm cm}^  
{-2}\ {\rm s}^{-1}$ ($4.7\times 10^{-13}\ {\rm erg} \  
{\rm cm}^{-2}\ {\rm s}^{-1}$) with a   
corresponding mean count rate of $9.0 \times 10^{-3}\  
{\rm counts}\ {\rm s}^{-1}  
$.  The resulting light curve is shown (arbitrarily scaled) in Figure 2. 
 
\subsubsection{UVOT Observations} 
 
 The UVOT \citep{2005SSRv..120...95R} began observing  
13 hours after the trigger. 
The data were retrieved from the \emph{Swift} archive at GSFC. 
Data taken in the uvm2 filter had not all been 
aspect-corrected, 
but inspection of the individual images showed that the sources fell 
within their 3\arcsec $\ $ radius region files.
 The counts were measured twice,  
using XIMAGE and UVOTMAGHIST, the 
latter during testing of the software updates for HEADAS version 6.3. The 
measurements were made for  3\arcsec $\ $ and 5\arcsec $\ $ radius 
apertures, both for the 
transient as 
well as for four field stars. Following the recommendations in  
\cite{2008MNRAS.383..627P}, 
an aperture correction was derived based on the observations of field stars 
having count rates of 1--10 ct/s, sufficiently low that the point-spread  
function (PSF) is not considered 
to be affected by the UVOT coincidence loss. The calibration of  
\cite{2008MNRAS.383..627P} was used (CALDB update of 2007-07-11).   
 
Since the UVOT is a photon-counting  
instrument, subsequent images can be co-added to improve the 
S/N. This was done by writing software to co-add the 
counts until a desired value of S/N was found as well as optimizing the time 
resolution. We checked that there was no noticeable difference in 
co-adding counts 
before or after doing background subtraction or coincidence-loss 
corrections, as 
expected for the low count rates of the transient.  The measurement 
error of the UVOT 
is Binomial due to the finite number of frames in an observation, and 
the errors were calculated accordingly with the formula given by  
\cite{2008MNRAS.383..383K}. Magnitudes were derived using the new  
zeropoints of \cite{2008MNRAS.383..627P}. 
 
The UVOT images of GRB\,070125 were further reduced following the 
recipe outlined by \cite{2006PASP..118...37L}, which provided an independent 
photometric calibration to the $U$, $B$, and $V$ filters employing a small 
aperture (radius of $2.5''$) for photometry 
(thus no need to do aperture correction).   
Comparison between the two reductions indicates that the measurements  
are consistent with each other within the uncertainties. 
 
\subsection{Ground-Based Observations} 
 
\subsubsection{Ground-Based Afterglow Photometry} 
 
The GCN burst notice \citep{2007GCN..6024....1H} was sent out  
at 21:46:48 Jan 25, approximately 13 hours and 26 minutes after  
the burst had been detected by the IPN.  The SARA 0.9~m telescope on Kitt  
Peak began imaging of the  
field of GRB\,070125 in the $V$ band  
$\sim$ 19 hours and 15 minutes after the trigger time, and  
was closely followed by the 0.41~m PROMPT telescope on Cerro Tololo 
 (20 hours and 34 minutes) in $BVRI$ 
and the Bok 2.3~m telescope on Kitt Peak in $V$ (24 hours). 
An afterglow candidate was detected by  
\cite{2007GCN..6028....1C} in the Palomar 1.5~m images.  Further 
observations were obtained with TNT, EST, SOAR, MDM, MMT, Kuiper, Loiano,  
KAIT, PAIRITEL, TNG, HCT, and the LBT (see Table~\ref{tab:obs}). 
Observations around 4 days after the trigger were hampered by  
the full Moon, which may explain why many telescopes stopped observing  
at that time. 
 
The SARA (Southeastern Association for Research in Astronomy) 
0.9~m telescope is located on Kitt Peak. SARA observations were carried out  
for three days following the burst until the afterglow was no longer  
detectable \citep{2007GCN..6029....1U}. Observations with SARA were 
limited to the $V$ band using the SARA Apogee Alta U47 camera. 
 
The Bok 2.3~m telescope is located at Kitt Peak  
National Observatory (KPNO) and is operated by the University of Arizona 
Steward Observatory. Bok observations utilized the 90prime 
instrument \citep{2004SPIE.5492..787W} and provide $BVRI$ data.  
Supplemental observations  
were obtained by the Kuiper Mont4k Imager  on the Kuiper 1.54~m  
telescope (located on Mt. Bigelow in Arizona) in the $R$ band.   
 
PROMPT observations were carried out in $BVRI$ by one of the five PROMPT  
(Panchromatic Optical Monitoring and Polarimetry Telescopes)  
telescopes located on Cerro Tololo, Chile. PROMPT5 is a 0.41~m 
Ritchey-Chr\'etien telescope outfitted with a rapid-readout  
Apogee Alta U47+ camera \citep{2007GCN..6044....1H}. 
 
The MDM observations were taken with the 2.4~m and 1.3~m Hiltner telescopes 
on Kitt Peak, a SITe backside-illuminated CCD, and $VRI$ filters. 
 
Observations from the Loiano 1.52~m telescope of the Bologna Astronomical 
Observatory, located at Loiano (Italy), were taken in the Cousins $R$  
band equipped with BFOSC, 
a multi-purpose instrument for imaging and spectroscopy  
\citep{2007GCN..6047....1G}. 
 
The 0.8~m TNT telescope and 1~m EST telescope are located at 
Xinglong Observatory of the National Astronomical Observatories of 
China. Each telescope is equipped with a Princeton Instruments 
$1340 \times 1300$ pixel CCD. Observations were carried out in the $R$ 
and $V$ bands. Observational details can be found in \cite{Deng2006}. 
 
The 3.58~m Telescopio Nazionale Galileo (TNG) is located  
at La Palma in the Canary Islands (Spain). TNG  
was equipped with the spectrophotometer DOLoRES (Device Optimized for  
the LOw RESolution) operating in imaging mode with a scale of $0.275''$ 
pixel$^{-1}$. 
 
The KAIT (Katzman Automatic Imaging Telescope;  
\citealt{2001ASPC..246..121F}) 0.76~m data 
are unfiltered, and were obtained with an Apogee AP7 camera.  
\cite{ 2003PASP..115..844L} 
demonstrated that the KAIT unfiltered magnitudes can be reliably transformed 
to the standard Cousins $R$ band with a precision of $\sim$5\%, if 
the color of the object is known. Since we have reliable color information for 
GRB\,070125 during the time of the KAIT observations, we calibrated the KAIT 
data to the $R$ band following the procedure in \cite{2003PASP..115..844L}.  
 
The SOAR (Southern Observatory for Astrophysical Research) 4.1~m telescope is  
located on Cerro Pachon, Chile. SOAR observations were taken in the near-IR  
using OSIRIS (Ohio State InfraRed Imager/Spectrometer). 
 
PAIRITEL (Peters Automated Infrared Imaging Telescope) is a 1.3~m  
telescope located on Mt. Hopkins in Arizona.  It is a robotic telescope 
which allows for rapid follow-up IR imaging of GRB targets.  The PAIRITEL observations 
were taken in the near-IR using a 2MASS instrument. 
 
The MMT  is located at the Whipple  
Observatory on Mt. Hopkins in Arizona. It is a 6.5~m telescope which 
contributed near-IR observations, obtained with SWIRC, the SAO 
Widefield Infrared Camera. 
  
Optical observations of the afterglow were carried out by the 2.01~m  
Himalayan Chandra Telescope (HCT) at the Indian Astronomical  
Observatory (IAO), Hanle (India). The CCD used at HCT was a  
2048 $\times$ 4096 pixel SITe chip mounted on the Himalayan Faint Object  
Spectrograph Camera (HFOSC).  Filters used are Bessell $V$, $R$ and $I$. 
 
The images were reduced and stacked where appropriate. Optical calibration was  
performed using IRAF PSF photometry and comparison to  
15 standard stars, whose magnitudes were obtained through a field calibration 
using Graham standard stars \citep{1982PASP...94..244G} and the Hardie method 
 \citep{1964aste.book.....H}. 
Near-IR images were reduced in IRAF using standard 
reduction pipelines. The MMT images were calibrated 
relative to standards observed during the same night, and the SOAR images 
were calibrated relative to the Graham standard stars.  LBT  
(Large Binocular Telescope) data cited are derived from 
\cite{2007GCN..6165....1G} and \cite{2007arXiv0712.2239D}.  Note that 
the Sloan $r'$ observation has been converted to the $R$ band.  All  
observations carried out by this collaboration can be found  
in Table~\ref{tab:obs}. 
 
\subsubsection{Ground-Based Afterglow Spectroscopy}  
 
We observed the afterglow of GRB~070125 with the Keck-I LRIS 
double spectrograph \citep{1995PASP..107..375O} through a long 
slit $1.0''$ wide starting 
at 05:44:05 on January 26. Two 600~s exposures were acquired using the 
D560 dichroic, the 400/3400 grism, and the 600/5000 grating. 
This instrumental configuration yields nearly continuous spectra over 
the range 3000--8000~{\AA}\ at a spectral resolution (full width 
at half-maximum intensity; FWHM) of $\sim$8~{\AA}\ on the blue side and  
$\sim$6~{\AA}\ on the red side. The data were reduced 
using standard procedures (bias subtracted, flat-fielded) and  
extracted with a boxcar encompassing $\sim$90\% of the flux. 

\subsubsection{Ground-Based Radio Observations} 
 
Radio observations were performed with the Westerbork Synthesis  
Radio Telescope (WSRT) at 1.4, 4.8, and 8.4 GHz. We used the Multi  
Frequency Front Ends \citep{1991ASPC...19...42T} in combination  
with the IVC+DZB back end\footnote{See \S 5.2 at  
http://www.astron.nl/wsrt/wsrtGuide/node6.html} in continuum mode,  
with a bandwidth of $8 \times 20$ MHz. Gain and phase calibrations  
were performed with the calibrator 3C~286. The first observation  
at 4.8~GHz, 1.5 days after the burst, was reported as a  
non-detection with a 3$\sigma$ upper limit of 261 $\mu$Jy  
\citep{2007GCN..6042....1V}, but a careful reanalysis of the data  
resulted in a 3.7$\sigma$ detection. The radio afterglow detection  
at $\sim5$ days was reported by \citet{2007GCN..6061....1C} and  
\citet{2007GCN..6063....1V}, using the Very Large Array at 8.64~GHz  
and the WSRT at 4.8~GHz respectively. The afterglow was bright  
enough to be detected up to 170 days after the burst at 4.8 GHz; after  
that, the flux dropped below the sensitivity limit. The measurement  
at 1.4 GHz resulted in a non-detection, while at 8.4 GHz we had a  
clear detection at 95 days after the burst. The details of our  
observations are shown in Table~\ref{tab:obs}; the 4.8~GHz light  
curve is shown in Figure~\ref{wsrt6cmresults}. 
 
\section{Data Analysis and Results} 
 
\subsection{The Light Curve} 
 
The light curve (Figure~\ref{fig:lc}) combines our optical,  
near-IR, and \emph{Swift} data. The bands have been offset from  
their actual magnitudes for ease in reading. The original  
unextinguished magnitudes 
are listed in Table~\ref{tab:obs}.   
Throughout this paper, we assume a standard flat cold dark matter cosmology 
($\Lambda$CDM), with parameters ($\Omega_{\Lambda}, \Omega_{\rm M}, H_{0}) =  
(0.761, 0.239, 73\ {\rm km}\  {\rm s}^{-1}  {\rm Mpc}^{-1})$,  
as found in the third-year WMAP data release \citep{2007ApJS..170..377S}  
assuming large-scale structure traced by luminous red  
galaxies (Tegmark et al. 2006). The particular set of values corresponds  
to the \textquotedblleft Vanilla model\textquotedblright of  
\cite{2006PhRvD..74l3507T}. 
 
\subsection{Fitting the Light Curve} 
 
\emph{Swift} observations began 0.54 days after the burst, followed  
by ground-based observations beginning 0.8 days by the SARA telescope,  
closely followed by the rest of the various telescopes collected in this  
collaborative effort. Due to a lack of early coverage, it is not immediately  
clear whether or not there is a jet break. We fit a broken power-law decay  
to each band separately using the Beuermann function  
\citep{1999A&A...352L..26B}  
as revised by  \cite{2001ApJ...546..117R}: 
 
\begin{equation} 
\noindent F_{\nu}(t)=2^{\frac{1}{n}} F_{\nu}(t_{b}) \left[ \left( \frac{t}{t_{b}} \right)^{\alpha_{1}n} 
+ \left( \frac{t}{t_{b}} \right)^{\alpha_{2}n} \right]^{-\frac{1}{n}}, 
\end{equation} 
 
\noindent where $F$ is the flux density in band $\nu$,  
$F_{\nu}(tb)$ is the flux  
density in band $\nu$ at the break time, $\alpha_{1}$ and $\alpha_{2}$ are 
the slopes of the power-law decay (before and after the jet break, 
respectively), and $n$ is the smoothness of the curve at the break. 
 
The complicated shape of our light curve (see Figure~\ref{fig:noflare}) 
makes fitting the overall structure difficult.  By using 
 the flux at 2 days, we shifted all bands to the extinction-corrected $R$ band 
for comparison and fitting.  The entire data set is obviously  
not well-fit by a broken power-law function.  Rapid  
flaring observed between one and two days post-trigger ($\S$ 3.5), 
influences the fit.  Broken power-law fits to the optical/UV/near-IR 
data set and the X-ray data set showed that this was indeed a poor 
model to the data set.  In addition, this fit predicted that  
the afterglow would have been nearly a magnitude brighter than  
it was observed to be at 26.8 days.  This would suggest the need  
for a second, non-standard jet break after 4 days. 
 
By eliminating the flaring region between 1 and 2 days  
after the trigger, we can better constrain our fit.  However, 
the best fit was determined by only considering the data taken after 
2 days.  At this point, it appears as if the flaring has settled 
to the point where the best estimate for the break time can be made. 
From the optical/UV/near-IR data set after 2 days, we  
determine an $\alpha_1$ = 1.56 $\pm$ 0.27, $\alpha_2$ = 2.47 
$\pm$ 0.13, and a jet break time of $t$ = 3.73 $\pm$ 0.52 days. 
While the X-ray data do not require a break before 9 days,
they are also not inconsistent with a break (see Table~\ref{tab:lctable}).
Lack of an achromatic break is not an uncommon feature in 
Swift-era bursts (e.g., \cite{2007MNRAS.381L..65C}). 
The results of all of the fits to the optical/UV/near-IR and  
X-ray data can be found in Table~\ref{tab:lctable}. 
 
Late-time MDM, LBT, and Chandra observations  
\citep{2007GCN..6186....1C} were invaluable  
in the determination 
of a late jet break time, as were the numerous early observations 
which allowed us to resolve the early flaring activity so as not 
to be confused with a jet break.  While the late MDM upper limit 
implied an optical jet break, the LBT detection at 26.8 days 
post-trigger confirmed the existence of a jet break.  Second-epoch 
deep imaging of the field of GRB\,070125 by the LBT did not  
result in the further detection of the source, thus allowing us 
to interpret the first detection as the afterglow  
\citep{2007arXiv0712.2239D}. Late-time Chandra X-ray Telescope 
observations also imply a late jet break in the X-ray light 
curve.  The Chandra upper limit of 2$\times$10$^{-15}$ erg cm$^{-2}$ 
s$^{-1}$ \citep{2007GCN..6186....1C} at 39.76 days after 
the burst was obtained in
the 0.3 -- 10 keV energy range, and is thus directly comparable 
to the XRT observations.  Fits to the XRT data (Table~\ref{tab:lctable})
do not include this upper limit.

The fluence of the burst was measured independently by the Konus-Wind 
and RHESSI instruments. By extrapolating to the GRB rest-frame energy band  
of 1 keV -- 10 MeV, we infer an isotropic energy of 
$E_{iso,Konus}=(9.44^{+0.40} 
_{-0.41})\times10^{53}$ erg and $E_{iso,RHESSI}=(8.27\pm0.39)\times10^{53}$  
erg \citep{2007arXiv0710.4590B}. 
Assuming a jet break at $t$ = 3.73 days and using the equations described in  
\cite{1999ApJ...519L..17S}, including a circumburst density of $n$ = 3 
${\rm cm}^{-3}$ and a gamma-ray efficiency of $\eta_{\gamma}$ = 0.2  
\citep{2007arXiv0710.4590B}, 
we find a jet half-opening angle of  
$\theta_{j,Konus}$= 
$5.65^{+0.03}_{-0.03}$ degrees  or $\theta_{j,RHESSI}$ = $5.74 \pm 0.04$ 
 degrees.   
For a jet break time of 3.73 days, the corresponding collimated energy  
emission  
is $E_{\gamma,Konus} = 4.58^{+0.24}_{-0.25} \times10^{51}$ erg,  
or $E_{\gamma,RHESSI} = (4.15 \pm 0.25)\times10^{51}$ erg following 
\cite{2003ApJ...594..674B}. 

Comprehensive broadband and energetics modeling of GRB\,070125 
was presented by Chandra et al. (2008).  They derived a circumburst density
of n $\sim$ 50 cm$^{-3}$ from the kinetic energy and n = 15.7 cm$^{-3}$
from a broadband fit using the synchrotron model.  If we use n = 50
cm$^{-3}$ to calculate $E_{\gamma}$, we find $E_{\gamma,Konus} =
9.26^{+0.49}_{-0.50}\times10^{51}$ erg ($\theta_{j,Konus} = 8.03 \pm 0.04$ 
degrees); $E_{\gamma,RHESSI} = 
(8.39 \pm 0.50)\times10^{51}$ erg ($\theta_{j,RHESSI} = 8.16 \pm 0.05$ degrees).  Using the broadband fit n = 15.7
cm$^{-3}$, we derive $E_{\gamma,Konus} = 6.93^{0.37}_{0.38}\times10^{51}$ erg 
($\theta_{j,Konus} = 6.94 \pm 0.03$ degrees),
$E_{\gamma,RHESSI} = (6.28 \pm 0.37)\times10^{51}$ erg ($\theta_{j,RHESSI} = 7.07 \pm 0.04$ degrees).  The most plausible
value for the circumburst density comes from the broadband fit, so the 
energies derived from n = 15.7 will be quoted as the energy of the burst
in this paper.
 
At $z$ = 1.547, a jet break at 3.73 days  
(Figure~\ref{fig:noflare}) implies a rather large energy release  
(when placed 
in the context of the sample studied by \citealt{2003ApJ...594..674B}).   
Both the observations of  
Konus-Wind and RHESSI  
imply that GRB\,070125 is one of the most energetic bursts discovered to date  
\citep{2003ApJ...594..674B}.

\subsection{Spectral Energy Distribution} 
 
Due to the variable nature of the early-time light curve,  
special care was used in determining the SED. We used the  
method employed by \cite{2006ApJ...641..993K} for deriving the SED  
of  GRB 030329. Using the $R$-band light curve as a reference, the  
other optical, UV, and near-IR  
bands were shifted until they matched the $R$-band light  
curve, which assumes the achromaticity already demonstrated above.  
Since we find no evidence for color evolution, use of the method is  
reasonable. The SED is then determined from the different colors,  
such as $U-R$. The $R$-band value is arbitrary, and thus also the  
absolute flux density scale of the SED. Also, due to achromaticity,  
no specific time should be associated with the SED shown in  
Figure~\ref{fig:sed2}.  
 To model the intrinsic optical extinction, we  
use the three best-modeled  
extinction curves --- those of the Milky Way (MW), Large  
Magellanic Cloud (LMC), and Small Magellanic Cloud (SMC) 
as parametrized by \cite{1992ApJ...395..130P}. 
We fit the SED with the method described by \cite{2006ApJ...641..993K}.  
The results  
are given in Table~\ref{tab:sed}. We find a clear preference for the  
SMC dust model ($A_{V} = 0.11\pm0.04$ mag).  
This preference, as well as  
the intrinsic  
spectral slope $\beta$ and the small host-galaxy extinction $A_{V}$,  
are all typical  
for the afterglows of (long) GRBs \citep[e.g.,][]{2006ApJ...641..993K, 
2007arXiv0712.2186K,2007ApJ...661..787S, Schady2007}. 
 
We combined the optical, UV, and near-IR photometry at,  
or extrapolated to, 4.26 days post-trigger 
with an X-ray spectrum extracted such that the log midpoint coincided 
with that of the optical SED described above. The {\it Swift} XRT X-ray  
spectrum was extracted using the 
method stated in $\S$ 2.1.1 and with the same extraction regions. The UVOT 
fluxes were obtained using the conversions in \cite{2008MNRAS.383..627P}. 
We calculated the transmission through the Lyman-$\alpha$ forest for each 
optical and UV band \citep[e.g.,][]{1995ApJ...441...18M}  
adopting the spectral slope 
$\beta=0.58$ derived from the optical SED, and corrected  
for these factors: $U$ transmission = 0.996, 
uvw1 = 0.848, uvm2 = 0.633, and uvw2 = 0.539. Galactic  
absorption and extinction were fixed at 
4.8 $\times$ 10$^{20}$ cm$^{-2}$ \citep{1990ARA&A..28..215D}  
and $E(B-V) = 0.052$ mag 
\citep{1998ApJ...500..525S}, respectively. 
The SED was created in count space following the 
method given by \cite{2007ApJ...661..787S}, and having the 
advantage that no model for the X-ray data need be assumed {\it a priori}. 
We fit the SED using models consisting of an absorbed power law or  
absorbed broken 
power law with slopes free or tied to $\Gamma_1 = \Gamma_2 - 0.5$ as expected 
for a cooling break. X-ray absorption  
is modeled with the 
Xspec model {\it zphabs} and assuming solar metallicity.  
 
We find that a broken power law provides a significantly better fit 
to the continuum than a single power law. The 
difference in slope between the two segments of the power law when both  
are left to vary tends to that expected for a cooling break:  
$\Delta\Gamma = 0.5$. Fixing the 
difference in the power-law slopes to 0.5 confirms this.  
(When compared to the single power law, the improvement in  
fit is significant according to the F-test, with probabilities 
of MW:10$^{-6}$, LMC:10$^{-9}$, and SMC:10$^{-8}$.)  
The cooling break can be 
well constrained to lie at 0.002 keV, 4.26 days since trigger.   
However, we caution that the distinction between the detection  
of a cooling break and a non-detection relies solely on the observed 
$K$-band flux, for which we have only two data points  
(see Table~\ref{tab:obs}).  Optical extinction is found to be 
$E(B-V) \approx 0.03$ mag (roughly equivalent to $A_V < 0.09$ mag),  
in full agreement with the fit from the optical SED alone.  
LMC and 
SMC extinction curves provide better fits than the MW extinction curve.  
The intrinsic X-ray absorption is relatively high, on the order of 
$N_{\rm H} \approx$ 2 $\times$ 10$^{21}$ cm$^{-2}$. Details of the  
fits are presented in  
Table~\ref{tab:sed2}. The unfolded SED and best-fitting model (BKNPL+LMC) are 
shown in Figure~\ref{fig:sed2}. 
 
\subsection{Light Curve and SED Analysis} 
 
The light curves at X-ray, UV, optical, and near-IR frequencies,  
and the SED spanning this large frequency range,  
can be utilized to determine the energy power-law index $p$ of  
the electrons emitting the synchrotron radiation.  
Adopting the standard blast-wave model, the temporal and spectral indices,  
$\alpha$ and $\beta$ (respectively), can be expressed in terms of 
 $p$ \citep{2004IJMPA..19.2385Z},  
assuming either a homogeneous or stellar wind-like circumburst medium.  
A single power-law fit to the X-ray light curve after the flare  
at $1.5 \times 10^5$  
seconds results in a temporal index of $\alpha_X=1.68^{+0.19}_{-0.15}$.  
For the analysis of the optical temporal slopes we take the fits without the  
flares (see Table~\ref{tab:lctable}), i.e., $\alpha_{\rm opt}=1.56\pm 0.27$  
before the break and  
$\alpha_{\rm opt}=2.47\pm 0.13$ after the break. 
 
The SED fits suggest that there is a cooling break at 0.002 keV (i.e. in the  
$R$ band) 4.26 days after trigger.  
However, this result should be interpreted with caution, since there is no  
achromatic light-curve break across the optical bands observed  
and the $K$-band flux is the only one that discriminates between presence or  
absence of a cooling break in the optical regime.  
Furthermore, the value of $p$ one can derive from the spectral slope above  
the tentative cooling break is $2\beta+1=2.12\pm 0.02$,  
which is not in agreement with the values one obtains from the X-ray temporal  
slope, $p=(4\alpha_{\rm X}+2)/3=2.91^{+0.25}_{-0.20}$,  
and from the optical temporal slope in the case of a homogeneous circumburst  
medium, $p=(4\alpha_{\rm opt}+3)/3=3.08\pm 0.36$.  
If the circumburst medium is structured like a stellar wind,  
$p=(4\alpha_{\rm opt}+1)/3=2.41\pm 0.36$ from the optical temporal slope  
is in agreement with the spectral slope,  
but the X-ray temporal slope remains inconsistent. Such a discrepancy 
is not unprecedented 
(e.g., \cite{2007astro.ph..3538P}).  However, we can derive an 
interpretation which is 
consistent with both the spectral and temporal fits (see below).
 
If one adopts the single power-law fit to the SED, all of the observed 
bands are either in between, or all of them are above, the peak frequency  
$\nu_m$ and cooling frequency $\nu_c$.  One then finds  
that $p=2\beta+1=2.98\pm 0.02$ 
or $p=2\beta=1.98\pm 0.02$, respectively. 
 
If one assumes that $\nu_m<\nu_X<\nu_c$, $p$ is equal to  
$(4\alpha_{\rm X}+3)/3=3.24^{+0.25}_{-0.20}$ or  
$(4\alpha_{\rm X}X+1)/3=2.57^{+0.25}_{-0.20}$,  
for a homogeneous or wind-like medium, respectively; if $\nu_m<\nu_c<\nu_X$,  
$p=(4\alpha_{\rm X}X+2)/3=2.91^{+0.25}_{-0.20}$.  
Comparing these values for $p$ with the ones from the single power-law SED  
fit, it is clear that they are only consistent for  
$\nu_m<\nu_X<\nu_c$ and a homogeneous medium.  
 
The pre-break optical temporal slope results in values for $p$ of  
$(4\alpha_{\rm opt}+3)/3=3.08\pm 0.36$ ($\nu_m<\nu_O<\nu_c$ and homogeneous medium),  
$(4\alpha_{\rm opt}+1)/3=2.41\pm 0.36$ ($\nu_m<\nu_O<\nu_c$ and stellar-wind medium),  
and $(4\alpha_{\rm opt}+2)/3=2.75\pm 0.36$ ($\nu_m<\nu_c<\nu_O$).  
Comparing these values for $p$ with the values from the SED fit,  
again $\nu_m<\nu_O<\nu_c$ and a homogeneous medium,  
which is the preferred situation from the X-ray temporal analysis,  
is consistent.  
 
Concluding, the X-ray to infrared bands at $\sim$4 days are  
situated in between $\nu_m$ and $\nu_c$, with $p \approx 3$ and the  
circumburst medium is homogeneous. We note that the $p$-value we 
derive for GRB\,070125 lies at the high end 
of the observed distribution, and is similar to that measured for 
e.g. GRB\,980519 ($p$=$2.96^{+0.06}_{-0.08}$, \cite{2008ApJ...672..433S}). 
\cite{2008ApJ...672..433S} showed that the $p$-distribution for a 
subsample of 
{\it BeppoSAX} GRBs is inconsistent with a single value of $p$ for all 
bursts at the 3$\sigma$ level, and constrain the intrinsic width of the 
parent distribution of $p$-values to 0.03$<\sigma_{\rm scat}<$1.40 
(3$\sigma$). A similar result was obtained for a {\it Swift} GRB subsample 
by  \cite{2006MNRAS.371.1441S}.
If we adopt a mean observed $p$-value of 2.04, as measured for the 
BeppoSAX subsample, then $p$=3 lies within the expected observational 
scatter. 
 
\subsection{Rebrightening Episodes} 
 
From the light curve (Figure~\ref{fig:lc}), 
we find evidence for two rebrightening or flaring episodes.   
These occur at $t$ = 1.15 days (as observed 
by the Bok telescope, SARA, TNT, and KAIT) and $t$ = 1.36 days (as observed by TNT and  
UVOT). The episode at $t$ = 1.15 days is the best sampled, 
as illustrated by the Bok data in Figure~\ref{fig:lcbok}. From 
the $V$ band alone, we derive a change in magnitude of  
about 0.5 (corresponding 
to a 56\% increase in flux) over a time period  
of $\Delta T \approx 8000$~s (0.093  
days), or an average increase in flux density of 17.1 $\mu$Jy/hour. 
 
Rebrightening episodes are not an uncommon afterglow phenomenon,  
and have been  
attributed to a variety of causes, including density fluctuations in the  
external medium (clumps, turbulence, or wind-termination shock structures), 
energy injection episodes (refreshed shocks) from the catch-up of faster shells 
in the outflow, patchy shells leading to angular inhomogeneities, extended  
activity of the central engine, multi-component jets, or, in rare cases,  
microlensing; see the recent summary by \cite{2007MNRAS.380.1744N} for links  
to  
the original literature on these options. \cite{2007MNRAS.380.1744N} 
 reconsidered  
the density fluctuation model with detailed numerical simulations, and find  
that  
sharp rebrightening episodes   
with   $\Delta T/T <$ 1 cannot be explained within this  
model. This result is in conflict with earlier conclusions, and the authors  
argue  
that their treatment of the reverse shock and photon travel-time effects  
explain 
the different outcome of their study.  
 
The rebrightening episodes seen in GRB\,070125 resemble those observed in 
GRB 021004 (\citealt{2005A&A...443..841D, 2003BASI...31...19P,  
2003Natur.422..284F,  
2003ApJ...584L..43B, 2003PASJ...55L..31U}),  
where three episodes with $\Delta T/T <$ 1 are noted at $t \approx 0.05$, 
0.8, and 2.6 days. The preferred interpretation for these episodes is  
fluctuations  
in the energy surface density, i.e., a patchy-shell model  
(\citealt{2003ApJ...594L..83G, 2004ApJ...602L..97N}).  
 
Another burst that lends itself to a comparison to GRB\,070125 is the  
well-sampled ``bumpy-ride'' event GRB 030329. Late-time energy 
injection (\textquotedblleft refreshed shocks\textquotedblright)  
has been invoked to explain the ``bumps'' in this  
GRB \citep{2003Natur.426..138G}. The well-resolved  
rebrightening episode discussed 
here is thus not likely caused by density fluctuations in the circumburst  
medium, but more likely attributed to angular or temporal energy fluctuations.  
Recently, \cite{2006ApJ...647.1238J} presented a detailed analysis of 
relativistic fireballs with discrete or continuous energy injection, and  
showed that energy injection imprints significant features  
on the afterglow, and thus provides a 
valuable diagnostic tool to study GRB fireball physics beyond the  
single-explosion  
standard model. In particular, \cite{2006ApJ...647.1238J} find that refreshed  
shocks from a discrete injection episode at $t$ = 1.4  
days (with an energy twice that of the initial 
energy injection) may, if not delivered uniformly across the shock surface  
(i.e., if a patchy shell is assumed), explain the sharp  
bump at $t$ = 3.5 days observed in GRB 000301C.  A more detailed  
follow-up paper is planned to further explore the rebrightening 
episodes found in GRB\,070125. 
 
\subsection{Radio Light Curve} 
 
The WSRT light curve at 4.8 GHz (see Figure~\ref{wsrt6cmresults}) displays  
the typical radio 
afterglow light-curve characteristics (e.g., \citealt{1997Natur.389..261F,  
1998ApJ...494L..49S}). 
Except for the first measurement, all the observations were performed 
after the jet-break time we derived from the optical light curves. 
At early times, the radio bands are situated below the synchrotron 
self-absorption frequency $\nu_a$ and the peak frequency $\nu_m$. 
Since these two characteristic frequencies move to lower observing 
frequencies in time, the light curve first rises up to the jet break, 
after which the flux remains constant. 
The measurements at 1.5 and 5.6 days seem to deviate from this behavior, 
but this can be explained by the effect of radio  
scintillation \citep{1997NewA....2..449G}. 
The rise in flux after $\sim$~40 days is caused by the passage of $\nu_m$ 
through the observing band, and the turnover at $\sim$~90 days by the 
passage of $\nu_a$, 
after which the light curve declines quite steeply with a temporal index 
equal to the electron energy power-law index $p \approx 3$. More 
detailed modeling of the radio light curve will be presented in a 
follow-up paper.
 
\subsection{Spectroscopy} 
 
It is evident from Figure~\ref{fig:aftspec} that the spectrum 
taken on Jan 26, 2007 at 05:44:05  
is nearly featureless.   
We have convolved the spectrum 
with a Gaussian matched to the instrumental resolution,  
and derive a $4\sigma$ equivalent width limits of approximately 
10~{\AA}\ at 3100~{\AA}, 5~{\AA} at 3200~{\AA}, 3~{\AA} at 3400~{\AA},  
1.5~{\AA}\ at 4000~{\AA}, and 1~{\AA}\ redward of 6000~{\AA}. 
We identify only a few features at $4\sigma$ statistical  
significance\footnote{Note that systematic errors (e.g.,\ continuum  
fitting) imply that the true significance limit is approximately $3\sigma$.} 
as listed in Table~\ref{tab:ew}. 
 
\cite{2007arXiv0712.2828C} have reported the detection of a relatively weak 
\ion{Mg}{2} doublet ($W_{MgII} < 1~${\AA}) in their afterglow spectrum of  
GRB~070125, leading to an implied redshift of 1.547. 
Our spectra do not confirm this line measurement, but the reported 
equivalent width lies below our detection threshold. However, 
we identify a strong feature at 3947.1~{\AA}\ which corresponds 
to the expected position of the \ion{C}{4} doublet for $z=1.547$. 
C$^{+3}$ gas is frequently associated with 
\ion{Mg}{2} absorbers \citep{2000ApJ...543..577C} 
and also gas surrounding GRB host galaxies \cite[e.g.,][]{2003ApJ...595..935M}. 
Unfortunately, we lack the spectral resolution 
to resolve the \ion{C}{4} doublet, but the likelihood of a  
misidentification is very low given the paucity of absorption lines 
in our data set. Furthermore, we observe weak  
absorption at the expected position of the \ion{Si}{4} doublet  
for $z=1.547$ which is spectrally resolved 
(Figure~\ref{fig:aftspec}) but has less than $4\sigma$ 
statistical significance. Altogether, the spectrum provides strong evidence 
for a metal-line absorption system at $z=1.547$, especially given  
the independent report of \ion{Mg}{2} absorption by \cite{2007GCN..6071....1F}. 
In turn, we establish this redshift as the lowest possible  
value for GRB\,070125. 
 
Of principal interest to our study is whether this metal-line 
absorption system results from gas in the GRB host galaxy and 
therefore establishes the redshift of GRB~070125. There are two 
indisputable signatures that an absorption system corresponds 
to a GRB host galaxy. (1) One observes fine-structure levels of  
O$^0$, Si$^+$, or Fe$^+$. These transitions, which have never been detected 
in an intervening system with quasar absorption-line spectroscopy,  
are known to arise from indirect radiative pumping by the GRB afterglow 
\citep{2006GCN..4593....1P,2007A&A...468...83V}. 
(2) One positively identifies the \lya\ forest  
imprint of the intergalactic medium and associates the highest 
redshift \lya\ absorption with the GRB.  In this latter case, one frequently 
observes a damped \lya\ (DLA) profile\footnote{\ion{H}{1} column  
density $\mnhi > 2\sci{20}\cm{-2}$.} at the GRB redshift  
\citep{2006A&A...447..897J}. 
 
Our spectrum reveals neither of these two signatures. First, 
we note no significant absorption from even the resonance lines 
of O$^0$, Si$^+$, or Fe$^+$, and we place relatively stringent upper 
limits on their equivalent widths  
(Table~\ref{tab:ew}). Second, our spectrum does not  
unambiguously show the \lya\ forest. Figure~\ref{fig:lya} presents 
the spectral region at $\lambda < 3400$~{\AA}. Overplotted on the 
data (green line) is a DLA profile centered at $z=1.547$ assuming  
$\mnhi = 10^{20.3} \cm{-2}$. Although the data formally reject 
the presence of a DLA system at this redshift, we caution 
that systematic effects (wavelength calibration, continuum placement) 
are not sufficiently large at $\lambda < 3100$~{\AA}\ to rule out a 
DLA profile at $>99\%$~c.l. The data do, however, rule out a DLA profile  
at all wavelengths greater than 3140~{\AA}. 
 
The absence of a strong DLA system or an obvious series 
of \lya\ absorption lines (i.e., the \lya\ forest) places an 
upper limit on the redshift of GRB~070125.  A reasonable 
estimate is to designate the absorption line (presently unidentified) 
at $\lambda = 3375$~{\AA}\ as a \lya\ transition and set $z_{GRB} \le  
1.78$. The identification of this feature as \lya\ 
is not supported by any coincident metal-line absorption  
(e.g., \ion{C}{4}). Furthermore, the implied 
rest equivalent width, $W_{Ly\alpha} = 1.4~${\AA}, would be  
significantly lower than any thus far reported for gas surrounding 
a GRB. Nevertheless, we cannot rule out this 
interpretation altogether. 
Applying Occam's razor to our full set of observations, we 
associate the metal-line absorption system at $z=1.547$ 
with the host galaxy of GRB~070125. To date, every absorption 
system which has been unambiguously associated with a GRB host 
galaxy (i.e.,\ using the criteria above) has shown strong \lya,  
\ion{Mg}{2}, and \ion{C}{4} absorption. 
In fact, this afterglow spectrum would already  
represent the weakest \ion{Mg}{2} doublet ever reported, 
although weaker \ion{C}{4} equivalent widths exist \citep{2006astro.ph..8327S}. 
 
Whether or not we associate the $z=1.547$ metal-line system with 
GRB~070125, this afterglow spectrum is remarkable in several 
respects. First, it is evident that the \ion{H}{1} column density 
is low. Taking $z_{GRB} = 1.547$, we have argued that the data 
prefer $\mnhi < 10^{20.3} \cm{-2}$, which is a rare but not 
unprecedented result \citep{2005ApJ...633..317F, 
2005GCN..3971....1P,2006GCN..4593....1P}.   
If $z>1.547$, then the \lya\ transition has the lowest equivalent 
width reported for a GRB. Either way, the low \nhi\ value is 
remarkable given the expectation that the GRB progenitor arises 
in a high density, star-forming region. Similar cases are discussed  
in \cite{2007ApJ...660L.101W}.  Second, even taking $z_{GRB}= 
1.547$ the \ion{C}{4}, \ion{Mg}{2}, and other low-ion equivalent  
widths are among the lowest observed to date for gas surrounding a  
GRB. Unfortunately, it is difficult to estimate the gas metallicity 
without a meaningful constraint on the \nhi\ value. Adopting $\mnhi =  
10^{19} \cm{-2}$, the upper limit to the equivalent width of \ion{Mg}{2} 
$\lambda$2796 implies a Mg/H ratio less than 1/10 the solar value. 
We derive a similar value adopting the upper limit to the 
equivalent width of \ion{Si}{2}~1526 ($W^{rest}_{1526} < 0.5$~{\AA}) and  
assuming the metallicity/$W_{1526}$ relation of \cite{2008ApJ...672...59P}. 
 
Taking both the spectroscopy and SED results into consideration, 
the host-galaxy properties of GRB\,070125 are consistent with 
those found in the sample presented by \cite{2006Natur.441..463F}. 
 
\section{Implications and Conclusions} 
 
\begin{figure} 
\plotone{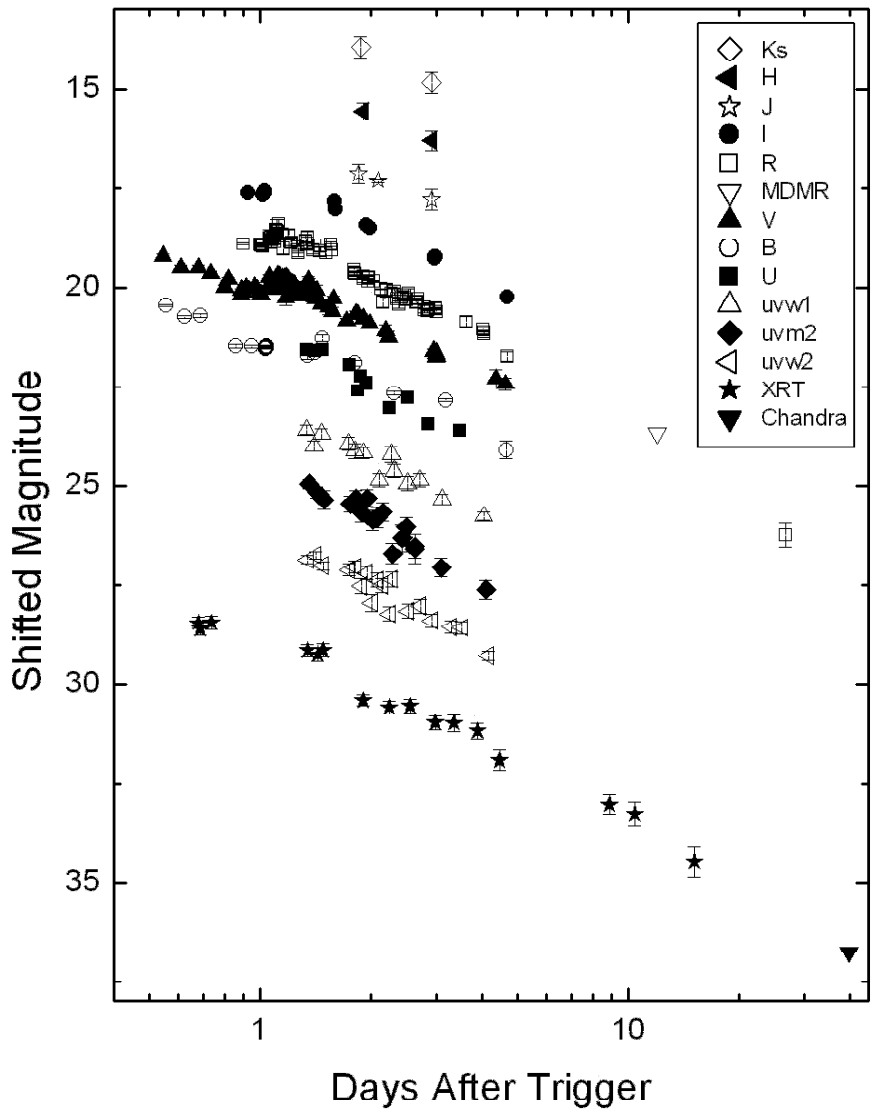} 
\caption{The combined data set (Vega magnitude system), corrected for Galactic extinction, arbitrarily 
scaled with respect to the $R$ band for ease in reading.  The XRT fluxes have 
been converted to magnitudes ($m = -2.5 \times {\rm log} F_{\nu}$). MDM and  
Chandra upper limits are represented by upside-down triangles.  Band magnitude shifts are $Ks -$ 3.0, $H -$ 2.0, $J -$ 1.0, $I -$ 0.5, $V +$ 1.0, $B +$ 2.0, $U +$ 3.0, uvw1 $+$ 5.0, uvm2 $+$ 6.5, uvw2 $+$ 8.0.  All errors (excluding upper limits) are shown.} 
\label{fig:lc} 
\end{figure}

\begin{figure}[ht] 
\begin{center} 
\plotone{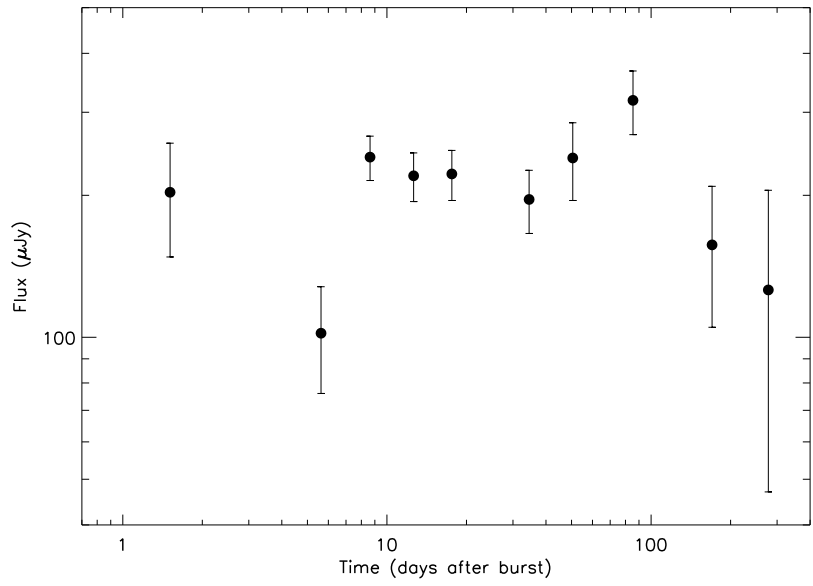} 
\caption{WSRT observations of GRB\,070125 at 4.8 GHz. 
\label{wsrt6cmresults}} 
\end{center} 
\end{figure}

\begin{figure} 
\begin{center} 
\plotone{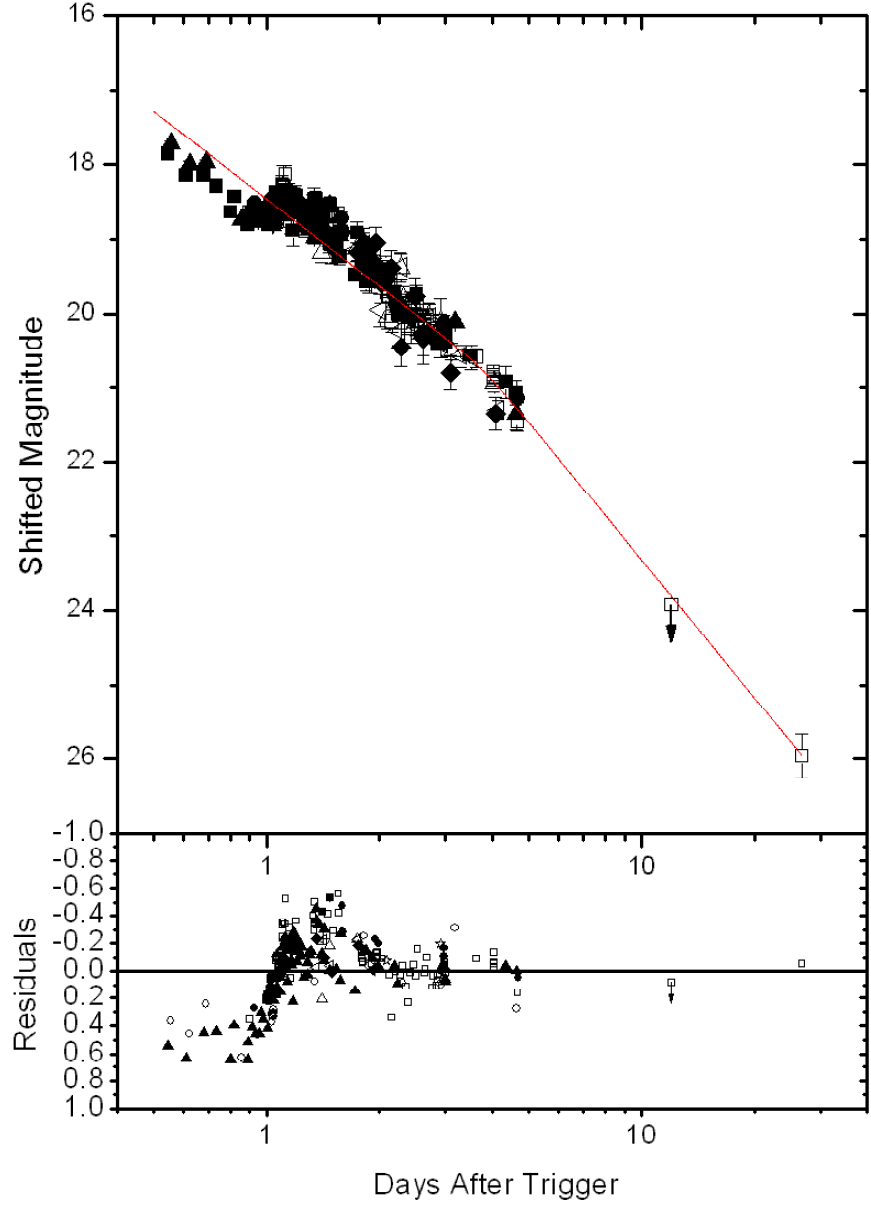} 
\caption{All UV, optical, and near-IR data contained in Figure~\ref{fig:lc} have been shifted to the $R$ band based on the flux at 2 days post-trigger.   Data before 2 days have been removed from this fit due to rapid flaring. The residuals to the fit are shown.  We derive a jet break time of 3.73 $\pm$ 0.52 days.} 
\label{fig:noflare} 
\end{center} 
\end{figure}

\begin{figure} 
\epsscale{0.8}
\plotone{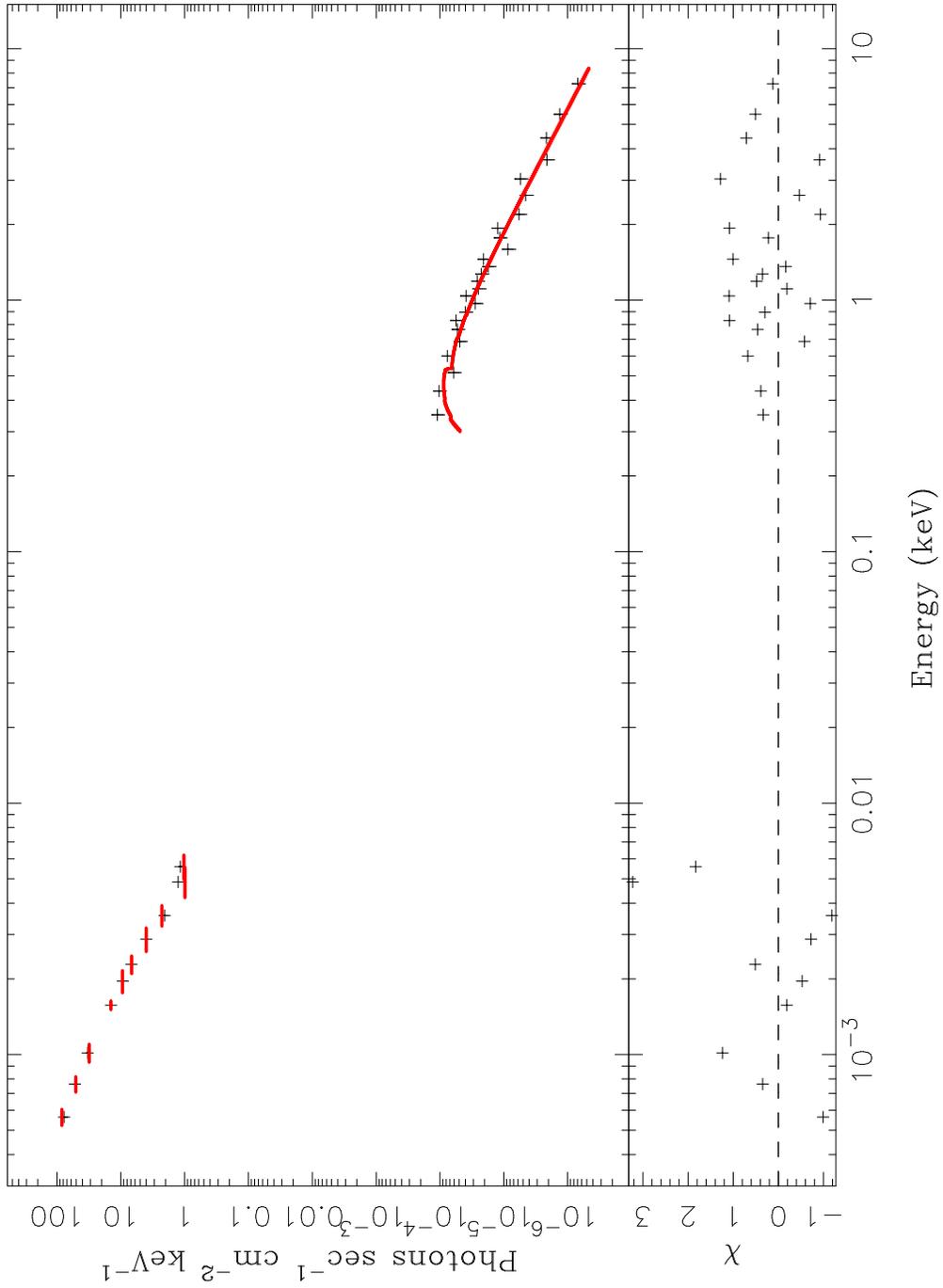} 
\caption{The unfolded SED at 4.26 days since trigger, including a {\it Swift} 
XRT spectrum and near-IR/optical/UV photometry (black points, upper panel). The 
 line indicates the best-fitting model of an absorbed, broken power law plus 
LMC-like extinction. The lower panel shows the residuals of the fit.} 
\label{fig:sed2} 
\end{figure}

\begin{figure} 
\begin{center} 
\plotone{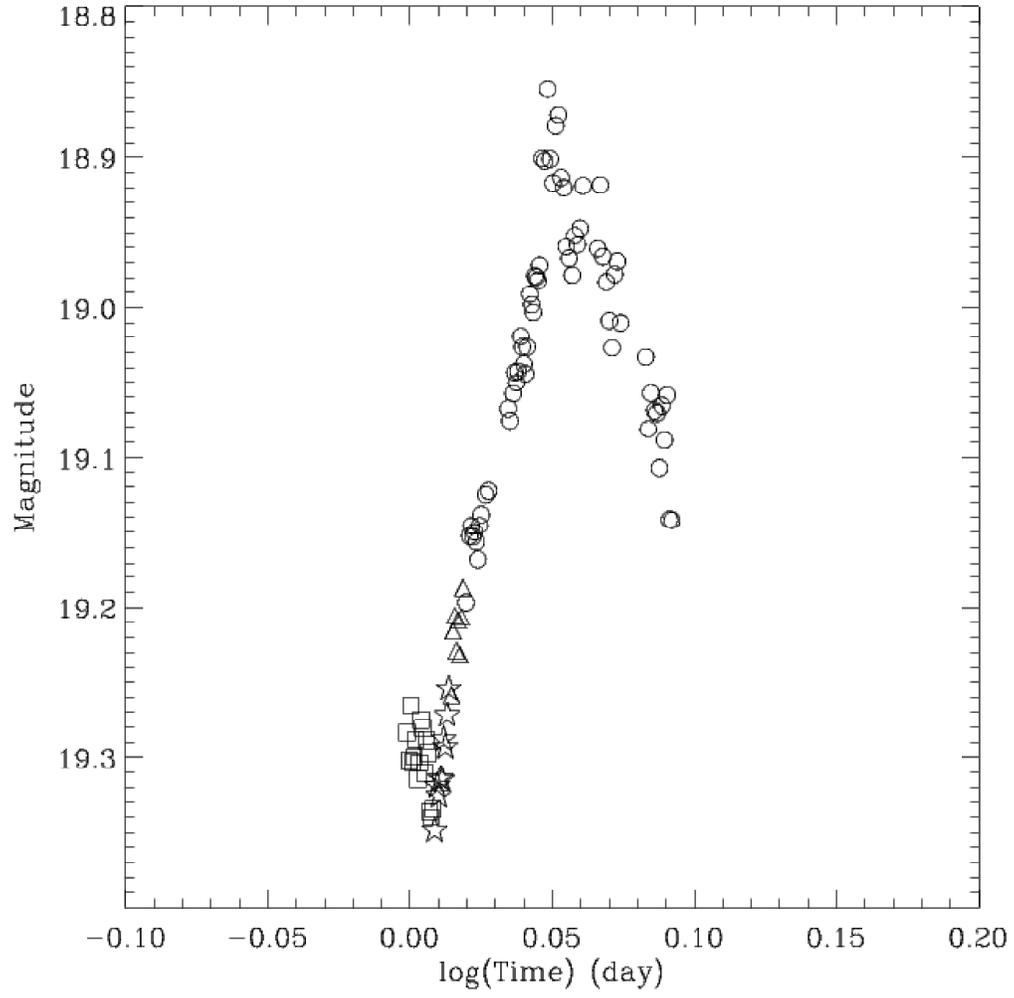} 
\caption{Bok $BVRI$ data shifted to the $V$ band, showing high S/N detection of 
the second rebrightening episode, which has been confirmed by KAIT, TNT, and 
SARA data.} 
\label{fig:lcbok} 
\end{center} 
\end{figure}

\begin{figure} 
\epsscale{0.8} 
\plotone{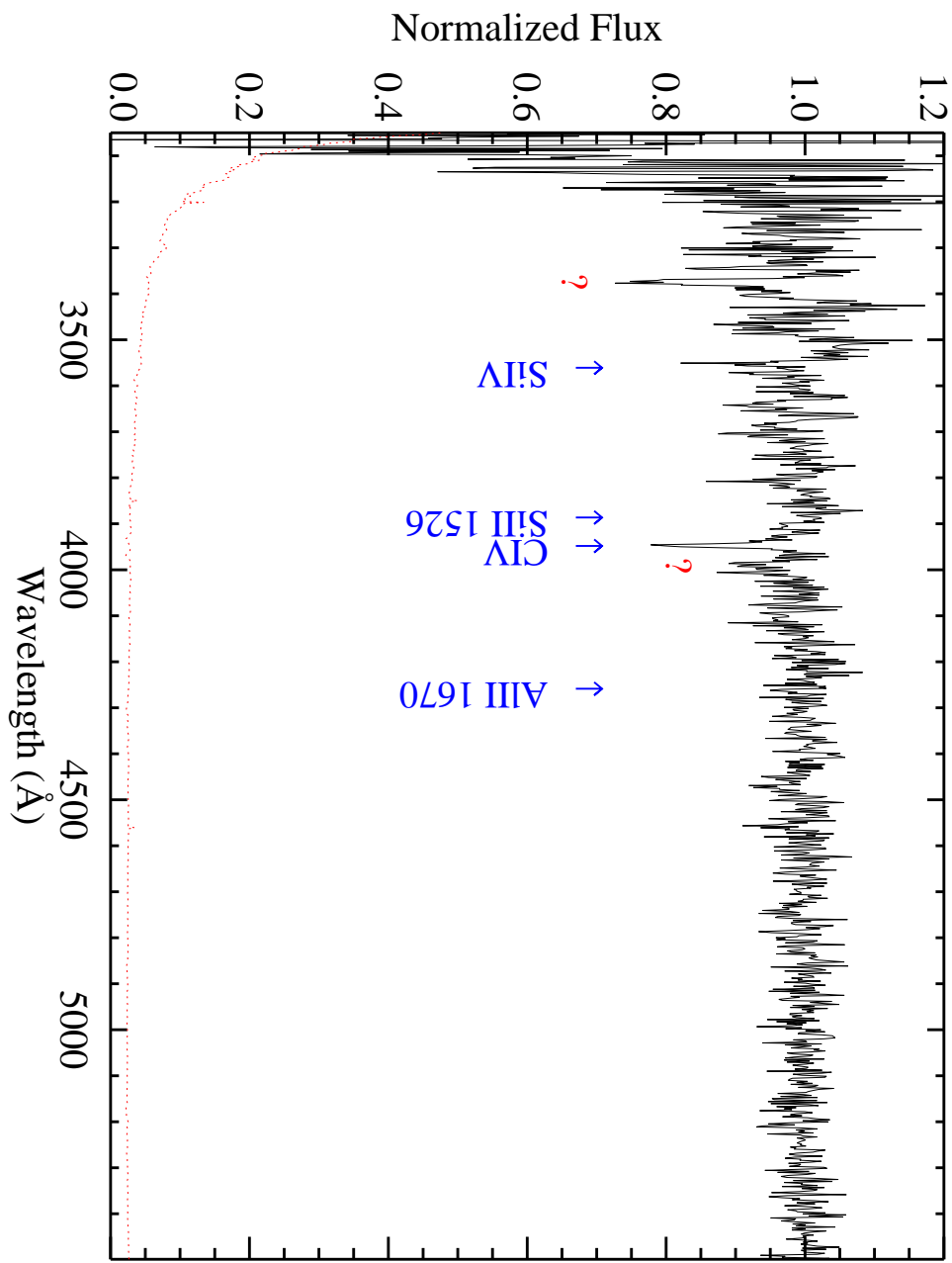} 
\caption{Normalized Keck/LRIS spectrum of the afterglow of GRB~070125. 
The figure shows the data acquired with the blue camera using 
the 400/3400 grism. 
The labeled arrows identify transitions associated with the metal 
absorption system at $z=1.547$. This gas most likely arises in the 
host galaxy of GRB~070125. We also mark several significant absorption  
features which remain unidentified.} 
\label{fig:aftspec} 
\end{figure}

\begin{figure} 
\epsscale{0.8} 
\plotone{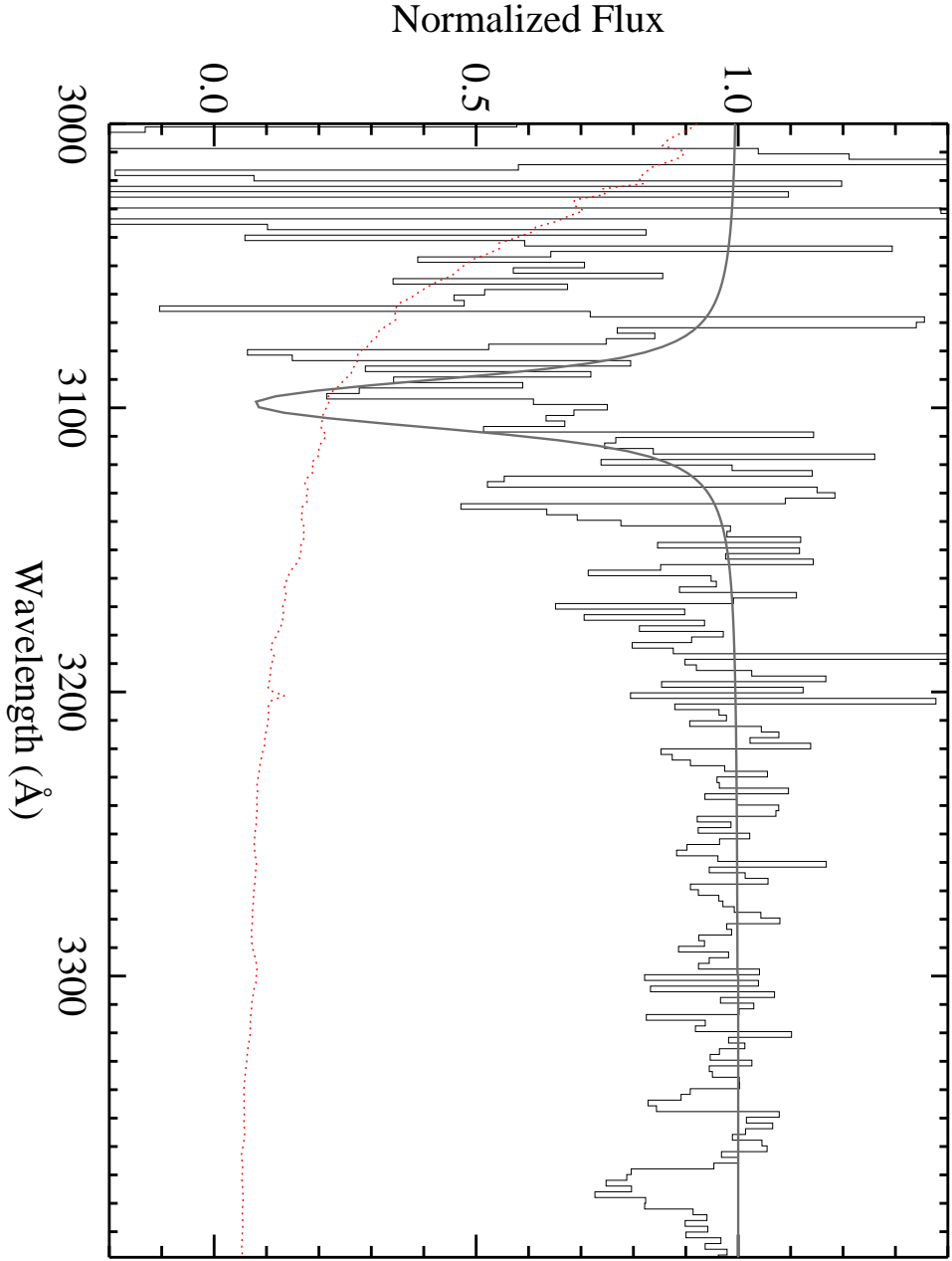} 
\caption{Close-up view of the spectral region near 3100~\AA\  
for the normalized Keck/LRIS spectrum of the GRB~070125 afterglow. 
The dotted line indicates the $1\sigma$ statistical error. The solid 
curve traces a damped \lya\ profile with $\mnhi = 10^{20.3} \cm{-2}$ 
centered at the redshift of the observed metal-line absorption 
system ($z=1.547$; Figure~\ref{fig:aftspec}). 
The data are formally inconsistent with a damped \lya\  
feature at this redshift, but we caution that systematic uncertainty 
(wavelength calibration and continuum placement error) do not rule 
out such a profile at very high confidence ($>99\%$~c.l.). 
The data do rule out the presence of a strong \lya\ absorber 
($W_{Ly\alpha} > 5$~\AA) redward of 3140~\AA.   
Furthermore, there is no obvious signature of 
the \lya\ forest in this spectrum.   
For these reasons, we argue that the metal-line absorber at $z=1.547$ 
is gas associated with the host galaxy of GRB~070125. 
We cannot rule out, however,  
the possibility that the feature at $\lambda \approx 3375$~\AA\ 
is relatively weak \lya\ absorption at $z=1.78$. 
This sets a formal upper limit to the redshift of GRB~070125. 
} 
\label{fig:lya} 
\end{figure}

While we do not have data before 0.545 days (the first exposures were obtained  
in $V$ with the UVOT), sampling thereafter is reasonably dense in time  
(see Figure~\ref{fig:lc}), until about 4 days after $T_0$.  
These observations suggest a possible jet break in the 1--3 day window. 
However, the strong  
variability (rebrightening episodes) around $t \approx$ 1--2 days  
interferes with a clean  
detection of the jet break, and it is conceivable that the actual  
afterglow light  
curve is not best represented by a broken power law (the Beuermann-like  
profile)  
with superimposed ``flares,'' but instead by  
three or more power laws, interrupted and perhaps re-established  
by several periods of rebrightening. GRB 030329 provides an  
example of a GRB with a jet break at $t \approx 0.5$ days  
(\citealt{2003Natur.423..843U,  
2003Natur.423..844P}), followed by multiple rebrightening episodes during  
subsequent days (e.g. \citealt{2003Natur.426..157G, 2004ApJ...606..381L}). 
Given that the light curve of GRB\,070125 exhibits strong variability, 
a jet-break interpretation requires extreme care.  However, the sparse 
late-time data do suggest that a break has occurred.  Fitting the data 
past 2 days indicates that the actual jet break is more likely to  
be identified with the required change in slope at $t$ = 3.7 days.    
If there indeed was no very early break in the optical power-law decay  
(before 0.5 days)  
and this late break time established from  
our fits is to be identified with the jet break, the indicated 
jet half-opening angle  
of $6.94^\circ$ implies $E_{\gamma}=6.93\times10^{51}$ erg (from  
$E_{iso,Konus}$), and GRB\,070125 is  
consistent with the Ghirlanda relation \citep{2004ApJ...616..331G}.  
 
\begin{figure} 
\plotone{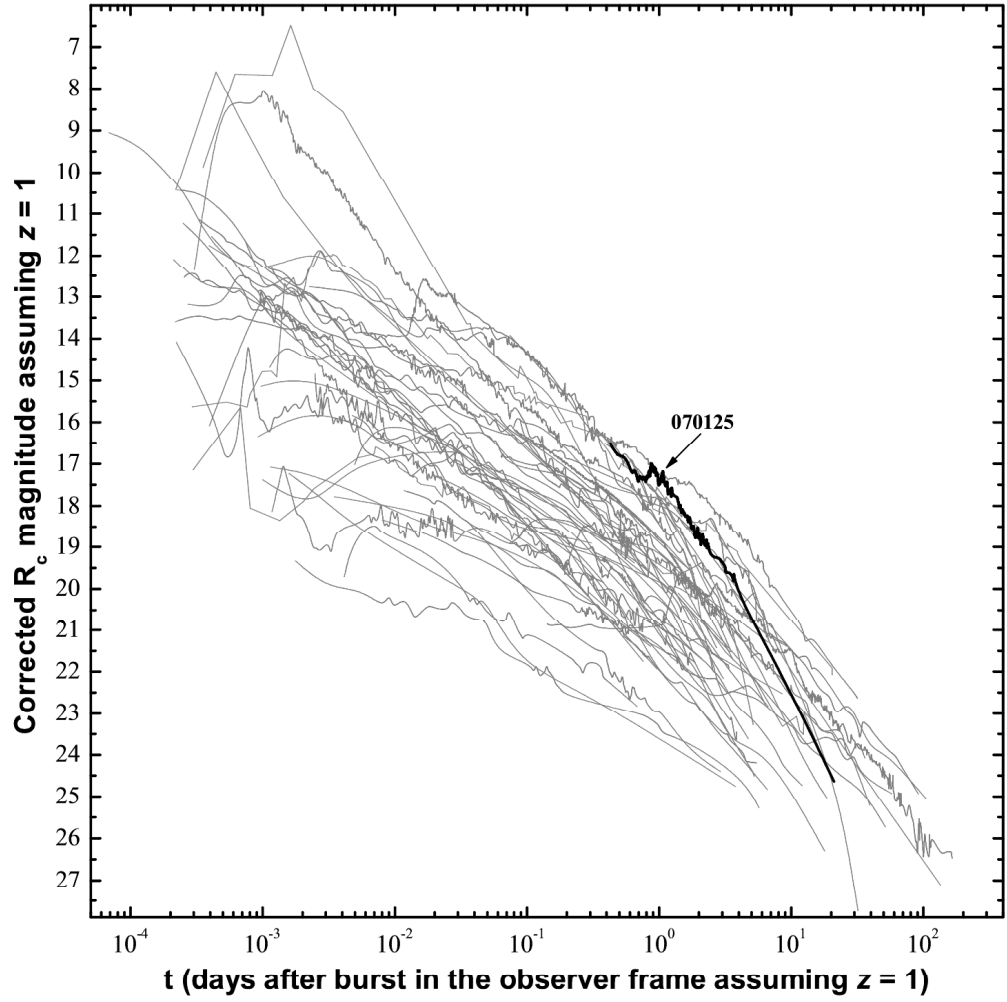} 
\caption{Afterglow light curves for 52 GRBs as they would look if all the 
bursts occurred at a redshift of 1 with no extinction 
\citep{2007arXiv0712.2186K}.  The thick line is the light curve of GRB\,070125.} 
\label{fig:big2} 
\end{figure}

The redshift allows us to place GRB\,070125 in general context.  
Figure~\ref{fig:big2} shows the intrinsic afterglow  
of GRB\,070125 in comparison with 52 other afterglows from the  
samples of \cite{2006ApJ...641..993K}  
and \cite{2007arXiv0712.2186K}. These afterglows are in the $R$  
band and have been corrected for 
Galactic extinction  
and, where possible, for host galaxy and supernova contributions.  
The afterglow of GRB\,070125  
was constructed by shifting $BVI$ data to the $R$ band, as only $BV$  
data are available at early times.  
Using the method described by \cite{2006ApJ...641..993K}, all  
afterglows have been corrected for rest-frame extinction and shifted  
to a common redshift of 1,  
allowing a direct comparison.  The afterglow of GRB\,070125  
is found to be among the most  
luminous afterglows; at the time of the rebrightening at one day,  
only the afterglow of GRB 021004  
is brighter.  The afterglow properties thus corroborate the  
inferences we made in the previous  
section, where the beaming-corrected burst energy in the 1--10,000  
keV regime \citep{2007GCN..6025....1B} 
was found to be on the high side of the observed statistical distribution. 
 
Had GRB\,070125 been much nearer, it would have given us  
an unprecedented opportunity to observe an  
afterglow on a much longer time scale, detect the  
additional emissions from an associated supernova, and  
possibly revealed a host galaxy. At a redshift of  
1.547, an unextinguished  
 SN 1998bw would have peaked at $R \approx  
27$ mag about one month after the burst. The LBT detection  
at late times could be explained with a supernova  
like SN 1998bw, if one allows a scaling factor of $k$ = 2.25, which  
is significantly larger than indicated by the properties  
of the sample of established GRB-supernovae  
(\citealt{2005NCimC..28..617Z, 2006A&A...457..857F}).  
Additionally, assuming that the late-time  
detection is composed entirely of supernova emission 
implies an even steeper afterglow decay.  Given the  
uncertainties when extrapolating the light-curve  
properties of SN 1998bw to a redshift of 1.5, we caution  that 
this factor has a large uncertainty. Based on what is known so far 
about the luminosities of GRB-SNe  
(\citealt{2004ApJ...609..952Z, 2006A&A...457..857F}),   
it seems unlikely that the LBT data point is indeed  
representing SN light.  Another alternative is light from an  
underlying host galaxy (not discernible in the images).   
Due to the second-epoch non-detection  
by the LBT \citep{2007arXiv0712.2239D},  
we can conclude that the host galaxy must be fainter than 
$M_V=18.2$ mag ($R > 26.1$ mag), which is comparable to   
the Magellanic Clouds or even fainter 
\citep{2000CAS....35.....V}, but  
consistent with the distribution found in the {\it Hubble Space Telescope} 
survey by \cite{2006Natur.441..463F}. The low metallicity  
we inferred from the spectra would also be  
consistent with a small, low-luminosity host galaxy.  
 
GRB\,070125 was unusual in several aspects. It had  
at least two sharp rebrightening episodes, it was 
intrinsically bright, and did not reveal the often  
associated DLA and Mg~II features in its smooth 
spectrum. The collective efforts of a large group allowed  
us to learn much about this burst, but it 
also provides a good example of the all-too-common  
situation that the world resources are too scarce 
to follow every burst with the kind of intensity in  
time and bandpass that may be required to extract 
enough information about each burst. Bumps have to be  
resolved, light curves need to be followed over 
longer time scales with greater sampling frequency,  
SEDs need to be established at many epochs, the 
associated supernovae should be checked even 
if the redshift is large, and host-galaxy properties  
should be established. Global coverage is necessary so  
that bumps and jet breaks are not missed, especially 
if future mission, such as EXIST \citep{2006AIPC..836..631G},
may result in much higher burst rates.

\acknowledgments 
 
A.C.U. and D.H.H. would like to thank Martha Leake and Matt Wood 
for generously sharing their SARA time with us. 
J.R.T. would like to acknowledge support from NSF grant 
AST-0307413.  D.A.K. and S.K. were supported by DFG grant 
Kl 766/13-2.  J.P.H. acknowledges support from Swift Guest Investigator  
Grant NNG06GJ25G.  The work of P.W.A.R., D.N.B., and J.L.R. was  
sponsored at Penn State University by NASA contract NAS5-00136.   
Observations reported here were in part obtained at the MMT  
Observatory, a joint facility of the University of Arizona and the  
Smithsonian Institution. J.D., W.Z, and Y.Q. are supported 
by NSFC Grant No. 10673014.  E.M. and E.P. are members of the  
Collaborazione Italiana Burst Ottici (CIBO) and thank the staff  
astronomers of the TNG for the excellent support.  Effort by C.W.H.  
was supported by a grant from NASA's Planetary Astronomy Program.   
R.L.C.S. acknowledges support from STFC.  
K.C.H. is grateful for IPN support under NASA grants NNG06GE69G 
and NNX06AI36G, and under JPL Contract 128043.  The Westerbork 
Synthesis Radio Telescope is operated by ASTRON (Netherlands Foundation 
for Research in Astronomy) with support from the Netherlands Foundation 
for Scientific Research (NWO).  A.J.v.d.H. was supported by an appointment 
to the NASA Postdoctoral Program at NSSTC, administered by Oak Ridge 
Associated Universities through a contract with NASA. A.V.F.'s group 
at UC Berkeley is supported by NSF grant AST--0607485, NASA/Swift 
grant NNG06GI86G, the TABASGO Foundation, and the Sylvia and Jim 
Katzman Foundation. Some of the data presented herein were obtained at  
the W. M. Keck Observatory, which is operated as a scientific partnership 
among the California Institute of Technology, the University of California 
and NASA. The Observatory was made possible by the generous financial 
support of the W. M. Keck Foundation.


 
 
 
\begin{deluxetable}{ c  c  c  c  l } 
\tablewidth{0pc} 
\tablecaption{OBSERVATIONAL CAMPAIGN\label{tab:obs}} 
\tabletypesize{\footnotesize} 
\tablehead{\colhead{Time} & \colhead{Mag or Flux} & \colhead{Error} & 
\colhead{Instrument} & \colhead{Band} }  
\startdata 
1.517	& 203 & 5.5E-5  & WSRT  &   4.8 GHz \\ 
5.642	& 102 & 2.6E-5  & WSRT  &   4.8 GHz \\ 
8.634	& 241 & 2.6E-5  & WSRT  &   4.8 GHz \\ 
12.929	& 220  & 2.6E-5  & WSRT  &   4.8 GHz \\ 
13.927	& -25.0 & 6.3E-5  & WSRT  &   1.4 GHz  \\ 
17.916	& 222 & 2.7E-5  & WSRT  &   4.8 GHz  \\ 
34.854	& 196 & 3.0E-5    & WSRT  &   4.8 GHz \\ 
50.827	& 240  & 4.5E-5  & WSRT  &   4.8 GHz \\ 
85.730	& 318 & 4.9E-5  & WSRT  &   4.8 GHz  \\ 
95.703	& 450  & 1.0E-4 & WSRT  &   8.4 GHz  \\ 
170.499	& 157 & 5.2E-5  & WSRT  &   4.8 GHz  \\ 
278.204	& 126 & 7.9E-5  & WSRT  &   4.8 GHz  \\ 
     1.349  &     19.25  &      0.10  & UVOT & uvw2   \\ 
     1.416  &     19.14  &      0.10  & UVOT & uvw2   \\ 
     1.484  &     19.41  &      0.11  & UVOT & uvw2   \\ 
     1.749  &     19.51  &      0.15  & UVOT & uvw2   \\ 
     1.816  &     19.44  &      0.15  & UVOT & uvw2   \\ 
     1.883  &     19.92  &      0.18  & UVOT & uvw2   \\ 
     1.950  &     19.59  &      0.16  & UVOT & uvw2   \\ 
     2.017  &     20.33  &      0.21  & UVOT & uvw2   \\ 
     2.085  &     19.77  &      0.17  & UVOT & uvw2   \\ 
     2.149  &     19.90  &      0.18  & UVOT & uvw2   \\ 
     2.252  &     20.61  &      0.18  & UVOT & uvw2   \\ 
     2.285  &     19.75  &      0.18  & UVOT & uvw2   \\ 
     2.521  &     20.54  &      0.17  & UVOT & uvw2   \\ 
     2.723  &     20.42  &      0.17  & UVOT & uvw2   \\ 
     2.929  &     20.79  &      0.18  & UVOT & uvw2   \\ 
     3.297  &     20.95  &      0.16  & UVOT & uvw2   \\ 
     3.525  &     20.97  &      0.17  & UVOT & uvw2   \\ 
     4.157  &     21.68  &      0.12  & UVOT & uvw2   \\ 
     1.360  &     18.91  &      0.11  & UVOT & uvm2   \\ 
     1.427  &     19.13  &      0.14  & UVOT & uvm2   \\ 
     1.492  &     19.31  &      0.24  & UVOT & uvm2   \\ 
     1.755  &     19.41  &      0.20  & UVOT & uvm2   \\ 
     1.822  &     19.28  &      0.20  & UVOT & uvm2   \\ 
     1.889  &     19.64  &      0.23  & UVOT & uvm2   \\ 
     1.956  &     19.27  &      0.21  & UVOT & uvm2   \\ 
     2.023  &     19.82  &      0.26  & UVOT & uvm2   \\ 
     2.090  &     19.76  &      0.25  & UVOT & uvm2   \\ 
     2.155  &     19.62  &      0.24  & UVOT & uvm2   \\ 
     2.290  &     20.68  &      0.26  & UVOT & uvm2   \\ 
     2.627  &     20.55  &      0.37  & UVOT & uvm2   \\ 
     3.101  &     21.03  &      0.23  & UVOT & uvm2   \\ 
     2.425  &     20.28  &      0.34  & UVOT & uvm2   \\ 
     2.491  &     20.00  &      0.25  & UVOT & uvm2   \\ 
     2.626  &     20.49  &      0.30  & UVOT & uvm2   \\ 
     4.098  &     21.58  &      0.23  & UVOT & uvm2   \\ 
     1.337  &     18.92  &      0.11  & UVOT & uvw1   \\ 
     1.404  &     19.33  &      0.13  & UVOT & uvw1   \\ 
     1.473  &     19.02  &      0.12  & UVOT & uvw1   \\ 
     1.743  &     19.27  &      0.17  & UVOT & uvw1   \\ 
     1.810  &     19.45  &      0.19  & UVOT & uvw1   \\ 
     1.910  &     19.50  &      0.14  & UVOT & uvw1  \\ 
     2.110  &     20.18  &      0.17  & UVOT & uvw1  \\ 
     2.313  &     19.95  &      0.17  & UVOT & uvw1  \\ 
     2.515  &     20.26  &      0.18  & UVOT & uvw1  \\ 
     2.716  &     20.18  &      0.17  & UVOT & uvw1  \\ 
     3.123  &     20.66  &      0.12  & UVOT & uvw1  \\ 
     2.279  &     19.53  &      0.20  & UVOT & uvw1  \\ 
     4.049  &     21.09  &      0.11  & UVOT & uvw1  \\ 
     1.340  &     18.81  &      0.10  & UVOT &    U  \\ 
     1.408  &     18.83  &      0.10  & UVOT &    U  \\ 
     1.476  &     18.81  &      0.10  & UVOT &    U  \\ 
     1.745  &     19.19  &      0.14  & UVOT &    U  \\ 
     1.845  &     19.85  &      0.13  & UVOT &    U  \\ 
     1.879  &     19.46  &      0.16  & UVOT &    U  \\ 
     1.946  &     19.65  &      0.17  & UVOT &    U  \\ 
     2.246  &     20.28  &      0.10  & UVOT &    U  \\ 
     2.517  &     20.03  &      0.13  & UVOT &    U  \\ 
     2.854  &     20.69  &      0.09  & UVOT &    U   \\ 
     3.487  &     20.85  &      0.08  & UVOT &    U  \\ 
0.555   &   18.66  &   0.03   &   UVOT       &   B    \\ 
0.622	 &   18.95  &	0.04   &   UVOT       &   B    \\ 
0.689	 &   18.92  &	0.04   &   UVOT       &   B    \\ 
0.857	 &   19.69  &	0.05   &   PROMPT     &   B    \\ 
0.947	 &   19.70   &	0.04   &   PROMPT     &   B    \\ 
1.034	 &   19.76  &	0.01   &   Bok        &   B    \\ 
1.035	 &   19.72  &	0.01   &   Bok	      &   B    \\ 
1.037	 &   19.71  &	0.01   &   Bok	      &   B    \\ 
1.038	 &   19.73  &	0.02   &   Bok	      &   B    \\ 
1.040	 &   19.71  &	0.01   &   Bok	      &   B    \\ 
1.041	 &   19.73  &	0.01   &   Bok	      &   B    \\ 
1.042	 &   19.71  &	0.01   &   Bok	      &   B    \\ 
1.044	 &   19.69  &	0.01   &   Bok	      &   B    \\ 
1.343	 &   19.92  &	0.09   &   UVOT       &   B    \\ 
1.410	 &   19.85  &	0.09   &   UVOT       &   B    \\ 
1.478	 &   19.49  &	0.08   &   UVOT       &   B    \\ 
1.813	 &   20.12  &	0.07   &   UVOT       &   B    \\ 
2.316	 &   20.89  &	0.05   &   UVOT       &   B    \\ 
3.187	 &   21.06  &	0.04   &   UVOT       &   B    \\ 
4.651	 &   22.31  &	0.22   &   TNG	      &   B    \\ 
0.545   &   18.37   &  0.05  & UVOT	     &  V      \\ 
0.611	 &   18.66   &  0.05  & UVOT	     &  V      \\ 
0.679	 &   18.66   &  0.05  & UVOT	     &  V      \\ 
0.736	 &   18.80    &  0.05  & UVOT	     &  V      \\ 
0.802	 &   19.15   &  0.07  & SARA	     &  V      \\ 
0.821	 &   18.95   &  0.04  & SARA	     &  V      \\ 
0.889	 &   19.33   &  0.05  & SARA	     &  V      \\ 
0.891	 &   19.21   &  0.02  & PROMPT      &   V     \\ 
0.914	 &   19.15   &  0.05  & SARA	     &  V      \\ 
0.935	 &   19.23   &  0.04  & SARA	     &  V      \\ 
0.957	 &   19.28   &  0.04  & SARA	     &  V      \\ 
0.967	 &   19.14   &  0.04  & PROMPT      &   V   	    \\ 
0.981	 &   19.21   &  0.04  & SARA	     &  V   	    \\ 
1.007	 &   19.33   &  0.03  & SARA	     &  V   	    \\ 
1.046	 &   19.20    &  0.01  & Bok 	     &  V   	    \\ 
1.049	 &   19.15   &  0.01  & Bok	     &  V   	    \\ 
1.051	 &   19.15   &  0.01  & Bok 	     &  V   	    \\ 
1.052	 &   19.15   &  0.01  & Bok 	     &  V   	    \\ 
1.054	 &   19.15   &  0.02  & Bok 	     &  V   	    \\ 
1.055	 &   19.16   &  0.01  & Bok 	     &  V   	    \\ 
1.056	 &   19.17   &  0.01  & Bok 	     &  V   	    \\ 
1.058	 &   19.15   &  0.02  & Bok 	     &  V   	    \\ 
1.059	 &   19.14   &  0.02  & Bok 	     &  V   	    \\ 
1.063	 &   18.88   &  0.05  & SARA	     &  V   	    \\ 
1.063	 &   19.13   &  0.01  & Bok 	     &  V   	    \\ 
1.066	 &   19.12   &  0.01  & Bok 	     &  V   	    \\ 
1.071	 &   19.06   &  0.04  & SARA	     &  V   	    \\ 
1.078	 &   19.05   &  0.04  & SARA	     &  V   	    \\ 
1.083	 &   19.07   &  0.01  & Bok	     &  V   	    \\ 
1.084	 &   19.08   &  0.01  & Bok	     &  V   	    \\ 
1.087	 &   19.06   &  0.01  & Bok	     &  V   	    \\ 
1.088	 &   19.20    &  0.05  & SARA	     &  V   	    \\ 
1.088	 &   19.04   &  0.01  & Bok	     &  V   	    \\ 
1.090	 &   19.05   &  0.01  & Bok	     &  V   	    \\ 
1.091	 &   19.04   &  0.01  & Bok	     &  V   	    \\ 
1.094	 &   19.02   &  0.01  & Bok	     &  V   	    \\ 
1.095	 &   19.03   &  0.01  & Bok	     &  V             \\ 
1.097	 &   19.04   &  0.01  & Bok	     &  V             \\ 
1.098	 &   19.04   &  0.01  & Bok	     &  V             \\ 
1.099	 &   18.98   &  0.06  & SARA	     &  V             \\ 
1.099	 &   19.03   &  0.01  & Bok	     &  V             \\ 
1.102	 &   18.99   &  0.02  & Bok	     &  V             \\ 
1.103	 &   19.00      &  0.01  & Bok	     &  V             \\ 
1.105	 &   19.00      &  0.01  & Bok	     &  V             \\ 
1.106	 &   18.98   &  0.01  & Bok	     &  V             \\ 
1.108	 &   19.04   &  0.05  & SARA	     &  V             \\ 
1.108	 &   18.98   &  0.01  & Bok	     &  V             \\ 
1.109	 &   18.98   &  0.02  & Bok	     &  V             \\ 
1.110	 &   18.97   &  0.01  & Bok	     &  V             \\ 
1.113	 &   18.90    &  0.03  & Bok	     &  V             \\ 
1.115	 &   18.90    &  0.03  & Bok	     &  V             \\ 
1.117	 &   19.05   &  0.06  & SARA	     &  V             \\ 
1.118	 &   18.85   &  0.04  & Bok	     &  V             \\ 
1.120	 &   18.90    &  0.03  & Bok	     &  V             \\ 
1.123	 &   18.92   &  0.03  & Bok	     &  V             \\ 
1.125	 &   18.88   &  0.03  & Bok	     &  V             \\ 
1.128	 &   18.87   &  0.03  & Bok	     &  V             \\ 
1.130	 &   18.91   &  0.03  & Bok	     &  V             \\ 
1.132	 &   18.92   &  0.03  & Bok	     &  V             \\ 
1.135	 &   18.96   &  0.02  & Bok	     &  V             \\ 
1.137	 &   19.19   &  0.05  & SARA	     &  V             \\ 
1.137	 &   18.97   &  0.03  & Bok	     &  V             \\ 
1.140	 &   18.98   &  0.01  & Bok	     &  V             \\ 
1.143	 &   18.95   &  0.02  & Bok	     &  V             \\ 
1.145	 &   18.96   &  0.03  & Bok	     &  V             \\ 
1.148	 &   18.95   &  0.04  & Bok	     &  V             \\ 
1.150	 &   18.92   &  0.03  & Bok	     &  V    	     \\ 
1.164	 &   19.08   &  0.02  & SARA	     &  V    	     \\ 
1.164	 &   18.96   &  0.04  & Bok	     &  V    	     \\ 
1.166	 &   19.13   &  0.20   & TNT 0.8m     &  V    	     \\ 
1.166	 &   18.92   &  0.02  & Bok	     &  V    	     \\ 
1.169	 &   18.97   &  0.04  & Bok	     &  V    	     \\ 
1.172	 &   18.98   &  0.02  & Bok	     &  V    	     \\ 
1.175	 &   19.01   &  0.02  & Bok	     &  V    	     \\ 
1.175	 &   19.40    &  0.20   & TNT 0.8m     &  V    	     \\ 
1.178	 &   19.03   &  0.02  & Bok	     &  V    	     \\ 
1.180	 &   18.91   &  0.03  & SARA	     &  V    	     \\ 
1.180	 &   18.98   &  0.02  & Bok	     &  V    	     \\ 
1.183	 &   18.97   &  0.02  & Bok	     &  V    	     \\ 
1.186	 &   18.98   &  0.15  & TNT 0.8m     &  V    	     \\ 
1.186	 &   19.01   &  0.03  & Bok	     &  V    	     \\ 
1.193	 &   18.94   &  0.04  & SARA	     &  V    	     \\ 
1.194	 &   19.07   &  0.15  & TNT 0.8m     &  V    	     \\ 
1.209	 &   19.16   &  0.05  & SARA	     &  V    	     \\ 
1.210	 &   19.03   &  0.02  & Bok	     &  V    	     \\ 
1.213	 &   19.08   &  0.01  & Bok	     &  V    	     \\ 
1.215	 &   19.06   &  0.01  & Bok	     &  V    	     \\ 
1.219	 &   19.07   &  0.01  & Bok	     &  V    	     \\ 
1.221	 &   19.07   &  0.02  & Bok	     &  V    	     \\ 
1.224	 &   19.11   &  0.02  & Bok	     &  V    	     \\ 
1.224	 &   19.06   &  0.04  & SARA	     &  V    	     \\ 
1.226	 &   19.07   &  0.02  & Bok 	     &  V    	     \\ 
1.227	 &   19.14   &  0.10   & TNT 0.8m     &  V    	     \\ 
1.229	 &   19.09   &  0.02  & Bok	     &  V    	     \\ 
1.231	 &   19.06   &  0.01  & Bok	     &  V    	     \\ 
1.234	 &   19.14   &  0.02  & Bok	     &  V    	     \\ 
1.236	 &   19.14   &  0.02  & Bok	     &  V             \\ 
1.241	 &   19.10    &  0.10   & TNT 0.8m     &  V             \\ 
1.241	 &   19.10    &  0.07  & EST 1m       &  V             \\ 
1.258	 &   19.35   &  0.12  & TNT 0.8m     &  V             \\ 
1.287	 &   19.28   &  0.12  & TNT 0.8m     &  V             \\ 
1.288	 &   19.37   &  0.10   & EST 1m       &  V             \\ 
1.299	 &   19.23   &  0.12  & TNT 0.8m     &  V             \\ 
1.312	 &   19.24   &  0.12  & TNT 0.8m     &  V             \\ 
1.355	 &   18.98   &  0.11  & UVOT	     &  V             \\ 
1.360	 &   19.07   &  0.14  & TNT 0.8m     &  V             \\ 
1.376	 &   19.12   &  0.12  & TNT 0.8m     &  V             \\ 
1.411	 &   19.38   &  0.12  & TNT 0.8m     &  V             \\ 
1.422	 &   19.44   &  0.12  & UVOT	     &  V             \\ 
1.423	 &   19.22   &  0.13  & TNT 0.8m     &  V             \\ 
1.472	 &   19.57   &  0.13  & TNT 0.8m     &  V             \\ 
1.536	 &   19.64   &  0.20   & TNT 0.8m     &  V             \\ 
1.572	 &   19.76   &  0.10   & HCT	     &  V             \\ 
1.582	 &   19.44   &  0.06  & HCT	     &  V             \\ 
1.721	 &   20.00      &  0.07  & UVOT	     &  V             \\ 
1.822	 &   19.81   &  0.06  & SARA        &   V   	    \\ 
1.848	 &   19.83   &  0.05  & PROMPT      &   V   	    \\ 
1.914	 &   19.93   &  0.05  & SARA        &   V   	    \\ 
1.987	 &   20.07   &  0.01  & MDM	     &  V   	    \\ 
2.199	 &   20.24   &  0.15  & TNT 0.8m     &  V   	    \\ 
2.237	 &   20.40    &  0.14  & TNT 0.8m     &  V   	    \\ 
2.964	 &   20.76   &  0.05  & SARA        &   V   	    \\ 
3.000	 &   20.88   &  0.02  & MDM	     &  V             \\ 
3.003	 &   20.81   &  0.02  & MDM	     &  V             \\ 
3.006	 &   20.91   &  0.03  & MDM	     &  V             \\ 
3.008	 &   20.89   &  0.02  & MDM	     &  V             \\ 
4.356	 &   21.45   &  0.22  & HCT	     &  V      \\ 
4.642	 &   21.60    &  0.16  & TNT 0.8m     &  V      \\ 
0.900   & 18.76   & 0.02	   & PROMPT    &  R   \\ 
0.998   & 18.78   & 0.02	   & Bok       &  R   \\ 
0.999   & 18.80    & 0.02	   & Bok       &  R   \\ 
1.001   & 18.77   & 0.02	   & Bok       &  R   \\ 
1.002   & 18.80    & 0.01	   & Bok       &  R   \\ 
1.003   & 18.80    & 0.01	   & Bok       &  R   \\ 
1.005   & 18.79   & 0.02	   & Bok       &  R   \\ 
1.006   & 18.81   & 0.02	   & Bok       &  R   \\ 
1.008   & 18.80    & 0.01	   & Bok       &  R   \\ 
1.009   & 18.78   & 0.01	   & Bok       &  R   \\ 
1.010   & 18.78   & 0.02	   & Bok       &  R   \\ 
1.019   & 18.81   & 0.02	   & Bok       &  R   \\ 
1.013   & 18.79   & 0.01	   & Bok       &  R   \\ 
1.015   & 18.80    & 0.02	   & Bok       &  R   \\ 
1.016   & 18.84   & 0.01	   & Bok       &  R   \\ 
1.017   & 18.84   & 0.02	   & Bok       &  R   \\ 
1.019   & 18.83   & 0.02	   & Bok       &  R   \\ 
1.063   & 18.61   & 0.21	   & KAIT      &  R   \\ 
1.065   & 18.60    & 0.16	   & KAIT      &  R   \\ 
1.070   & 18.75   & 0.11	   & KAIT      &  R   \\ 
1.070   & 18.59   & 0.11	   & KAIT      &  R   \\ 
1.075   & 18.67   & 0.09	   & KAIT      &  R   \\ 
1.078   & 18.66   & 0.07	   & KAIT      &  R   \\ 
1.081   & 18.68   & 0.07	   & KAIT      &  R   \\ 
1.084   & 18.63   & 0.07	   & KAIT      &  R   \\ 
1.087   & 18.60    & 0.08	   & KAIT      &  R   \\ 
1.087   & 18.66   & 0.08	   & KAIT      &  R   \\ 
1.093   & 18.59   & 0.08	   & KAIT      &  R   \\ 
1.095   & 18.63   & 0.08	   & KAIT      &  R   \\ 
1.095   & 18.71   & 0.08	   & KAIT      &  R   \\ 
1.102   & 18.40    & 0.13	   & KAIT      &  R   \\ 
1.104   & 18.42   & 0.10	   & KAIT      &  R   \\ 
1.104   & 18.51   & 0.14	   & KAIT      &  R   \\ 
1.110   & 18.41   & 0.27	   & KAIT      &  R   \\ 
1.120   & 18.44   & 0.11	   & KAIT      &  R   \\ 
1.120   & 18.58   & 0.15	   & KAIT      &  R   \\ 
1.125   & 18.26   & 0.12	   & KAIT      &  R   \\ 
1.151   & 18.52   & 0.14	   & TNT 0.8m  &  R   \\ 
1.159   & 18.88   & 0.17	   & TNT 0.8m  &  R   \\ 
1.201   & 18.53   & 0.13	   & TNT 0.8m  &  R   \\ 
1.209   & 18.76   & 0.10	   & TNT 0.8m  &  R   \\ 
1.222   & 18.71   & 0.10	   & EST 1m    &  R   \\ 
1.266   & 19.00   & 0.10	   & EST 1m    &  R   \\ 
1.273   & 18.85   & 0.09	   & TNT 0.8m  &  R   \\ 
1.331   & 18.68   & 0.15	   & TNT 0.8m  &  R   \\ 
1.339   & 18.79   & 0.15	   & TNT 0.8m  &  R   \\ 
1.343   & 18.58   & 0.15	   & TNT 0.8m  &  R   \\ 
1.346   & 18.74   & 0.16	   & TNT 0.8m  &  R   \\ 
1.394   & 18.93   & 0.10	   & TNT 0.8m  &  R   \\ 
1.450   & 18.96   & 0.10	   & TNT 0.8m  &  R   \\ 
1.462   & 18.82   & 0.17	   & TNT 0.8m  &  R   \\ 
1.513   & 19.00   & 0.14	   & TNT 0.8m  &  R   \\ 
1.555   & 18.77   & 0.06	   & HCT       &  R   \\ 
1.563   & 18.92   & 0.02	   & HCT       &  R   \\ 
1.796   & 19.43   & 0.05	   & Kuiper    &  R   \\ 
1.802   & 19.49   & 0.04	   & Kuiper    &  R   \\ 
1.807   & 19.53   & 0.05	   & Kuiper    &  R   \\ 
1.813   & 19.52   & 0.04	   & Kuiper    &  R   \\ 
1.905   & 19.58   & 0.03	   & PROMPT    &  R   \\ 
1.912   & 19.63   & 0.02	   & Kuiper    &  R   \\ 
1.964   & 19.60    & 0.04	   & MDM       &  R   \\ 
1.968   & 19.70    & 0.02	   & MDM       &  R  	\\ 
1.971   & 19.64   & 0.02	   & MDM       &  R  	\\ 
1.975   & 19.64   & 0.01	   & MDM       &  R  	\\ 
1.979   & 19.64   & 0.01	   & MDM       &  R  	\\ 
1.983   & 19.61   & 0.02	   & MDM       &  R  	\\ 
2.036   & 19.70    & 0.01	   & Kuiper    &  R  	\\ 
2.130   & 19.90    & 0.01	   & Kuiper    &  R  	\\ 
2.160   & 20.23   & 0.15	   & TNT 0.8m  &  R  	\\ 
2.213   & 19.93   & 0.18	   & EST 1m    &  R  	\\ 
2.275   & 20.00   & 0.11	   & TNT 0.8m  &  R  	\\ 
2.312   & 19.97   & 0.10	   & TNT 0.8m  &  R  	\\ 
2.350   & 20.15   & 0.10	   & TNT 0.8m  &  R  	\\ 
2.387   & 20.29   & 0.10	   & TNT 0.8m  &  R  	\\ 
2.427   & 20.10    & 0.10	   & TNT 0.8m  &  R  	\\ 
2.464   & 20.07   & 0.10	   & TNT 0.8m  &  R  	\\ 
2.500   & 20.17   & 0.12	   & TNT 0.8m  &  R  	\\ 
2.533   & 20.01   & 0.12	   & TNT 0.8m  &  R  	\\ 
2.644   & 20.25   & 0.04	   & Loiano    &  R  	\\ 
2.659   & 20.23   & 0.05	   & Loiano    &  R  	\\ 
2.677   & 20.16   & 0.04	   & Loiano    &  R  	\\ 
2.781   & 20.44   & 0.06	   & Loiano    &  R  	\\ 
2.795   & 20.36   & 0.04	   & Loiano    &  R  	\\ 
2.843   & 20.46   & 0.04	   & Kuiper    &  R  	\\ 
2.854   & 20.42   & 0.03	   & Kuiper    &  R  	\\ 
2.865   & 20.40    & 0.02	   & Kuiper    &  R  	\\ 
2.989   & 20.40    & 0.02	   & MDM       &  R  	\\ 
2.992   & 20.38   & 0.02	   & MDM       &  R  	\\ 
2.995   & 20.40    & 0.02	   & MDM       &  R  	\\ 
2.997   & 20.41   & 0.02	   & MDM       &  R  	\\ 
3.006   & 20.47   & 0.03	   & Kuiper    &  R  	\\ 
3.631   & 20.73   & 0.14	   & Loiano    &  R  	\\ 
4.034   & 20.92   & 0.04	   & MDM       &  R  	\\ 
4.038   & 21.00   & 0.03	   & MDM       &  R  	\\ 
4.042   & 20.99   & 0.05	   & MDM       &  R  	\\ 
4.046   & 21.03   & 0.05	   & MDM       &  R  	\\ 
4.678   & 20.97   & 0.12	   & TNG       &  R   \\ 
4.866   & 22.23   & 0.08	   & Kuiper    &  R    \\ 
12.001  & 23.80    & Upper limit   & MDM       &  R    \\ 
26.800  & 26.10    & 0.30	   & LBT       &  R    \\ 
0.929  & 18.21  & 0.02  & PROMPT    & I  \\ 
1.020	& 18.25  & 0.02  & Bok	     & I  \\ 
1.022	& 18.22  & 0.02  & Bok	     & I  \\ 
1.023	& 18.23  & 0.03  & Bok	     & I  \\ 
1.025	& 18.21  & 0.03  & Bok	     & I  \\ 
1.026	& 18.21  & 0.02  & Bok	     & I  \\ 
1.028	& 18.19  & 0.03  & Bok	     & I  \\ 
1.029	& 18.19  & 0.03  & Bok	     & I  \\ 
1.030	& 18.17  & 0.03  & Bok	     & I  \\ 
1.039	& 18.15  & 0.03  & Bok	     & I  \\ 
1.591	& 18.41  & 0.09  & HCT	     & I  \\ 
1.599	& 18.61  & 0.09  & HCT	     & I  \\ 
1.947	& 19.01  & 0.05  & PROMPT    & I  \\ 
1.992	& 19.08  & 0.02  & MDM	     & I  \\ 
2.977	& 19.88  & 0.02  & MDM	     & I  \\ 
2.980	& 19.82  & 0.02  & MDM	     & I  \\ 
2.982	& 19.82  & 0.03  & MDM	     & I  \\ 
4.693	& 20.84  & 0.06  & TNG	     & I  \\ 
1.854	   &  	   18.17	&     0.24  &    SOAR     &   J  \\ 
2.090  &  	   18.35	&     0.04  &    MMT	    &   J  \\ 
2.916    &       18.82	&     0.26  &    PAIRITEL &   J \\ 
1.907 &       17.58	&     0.22  &    SOAR     &   H  \\ 
2.916    &       18.33	&     0.25  &    PAIRITEL  &  H  \\  
1.877  &  	   16.96	&     0.28  &    SOAR     &  Ks  \\ 
2.916    &       17.86	&     0.25  &    PAIRITEL &  Ks\\   
0.550    &	    5.44e-12   &    4.95e-13   &    XRT	&   X   \\ 
0.610    &	    4.23e-12   &    6.47e-13   &    XRT	&   X   \\ 
0.621    &	    3.51e-12   &    5.30e-13   &    XRT	&   X   \\ 
0.679    &	    4.16e-12   &    6.36e-13   &    XRT	&   X   \\ 
0.688    &	    3.62e-12   &    5.70e-13   &    XRT	&   X   \\ 
0.737    &	    4.19e-12   &    7.12e-13   &    XRT	&   X   \\ 
1.349     &	    2.19e-12   &    3.41e-13   &    XRT	&   X   \\ 
1.437     &	    2.00e-12   &    3.23e-13   &    XRT	&   X   \\ 
1.483     &	    2.19e-12   &    3.61e-13   &    XRT	&   X   \\ 
1.909     &	    6.85e-13   &    9.24e-14   &    XRT	&   X   \\ 
2.248     &	    5.83e-13   &    9.04e-14   &    XRT	&   X   \\ 
2.552     &	    6.01e-13   &    1.09e-13   &    XRT	&   X   \\ 
2.988     &	    4.08e-13   &    7.35e-14   &    XRT	&   X   \\ 
3.359     &	    4.05e-13   &    8.70e-14   &    XRT	&   X   \\ 
3.896     &	    3.37e-13   &    6.75e-14   &    XRT	&   X   \\ 
4.470     &	    1.73e-13   &    4.86e-14   &    XRT	&   X   \\ 
8.884     &	    6.21e-14   &    1.55e-14   &    XRT	&   X   \\ 
10.423     &	    4.99e-14   &    1.55e-14   &    XRT	&   X   \\ 
15.139     &       1.62e-14   &    6.75e-15   &    XRT   &   X   \\  
39.760       &       2e-15         &    Upper limit     &    Chandra & X   \\ \hline			              		      	  	     
\enddata 		              		       
\tablecomments{Observations carried out by this collaboration, late-time LBT 
detection \citep{2007arXiv0712.2239D}, and late-time Chandra observation 
\citep{2007GCN..6186....1C}.  X-ray fluxes have units of ${\rm erg}\ {\rm 
cm}^{-2}\ {\rm s}^{-1}$.  Radio fluxes are given in ${\rm \mu Jy}$.} 
\end{deluxetable} 	               
			              
 
\begin{deluxetable}{ c  c  c  c  c  c } 
\tablewidth{0pc} 
\tablecaption{LIGHT-CURVE FITTING\label{tab:lctable}} 
 
\tabletypesize{\footnotesize} 
\tablehead{\colhead{Band} & \colhead{Model} & \colhead{$\alpha_1$} & \colhead{$\alpha_2$} & \colhead{Break Time} & \colhead{$\chi^{2}$/d.o.f.}} 
\startdata 
opt/uv/nir & all data & 0.47 $\pm$ 0.10 & 2.05 $\pm$ 0.04 & 1.39 $\pm$ 0.04 & 10.25 \\ 
opt/uv/nir & no flaring & 1.24 $\pm$ 0.09 & 2.49 $\pm$ 0.13 & 3.37 $\pm$ 0.36 & 3.35 \\ 
opt/uv/nir & after two days & 1.56 $\pm$ 0.27 & 2.47 $\pm$ 0.13 & 3.73 $\pm$ 0.52 & 2.98 \\  
X-ray & all data & $1.16^{+0.23}_{-0.87}$ & $1.85^{+0.19}_{-0.15}$ & $1.34^{+0.24}_{-0.51}$ & 1.6 \\ 
X-ray & no flaring & $1.49^{+0.14}_{-0.09}$ & $1.94^{+0.39}_{-0.30}$ & $3.93^{+9.72}_{-1.85}$ & 0.80 \\ 
X-ray & after two days & $1.0^{+0.62}_{-1.81}$ & $1.99^{+0.46}_{-0.32}$ & $3.34^{+9.03}_{-1.27}$ & 0.50 \\ \hline 
\enddata 		              		       
\tablecomments{Results of fitting a Beuermann function 
\citep{1999A&A...352L..26B} (as revised by  \citealt{2001ApJ...546..117R}) 
broken power-law function to the data set; 
opt/uv/nir refers to the combined data set of optical, UV, and  
near-IR data.  The  
columns labelled ``no flaring'' have removed the data between 1  
and 2 days from the fit.}\label{tab:lc} 
\end{deluxetable}

 
\begin{deluxetable}{ c  c  c  c } 
\tablewidth{0pc} 
\tablecaption{UV/OPTICAL/NIR SPECTRAL ENERGY DISTRIBUTION\label{tab:sed}} 
\tabletypesize{\footnotesize} 
\tablehead{\colhead{dust} & \colhead{$\chi^2$/d.o.f.} & \colhead{$\beta$} & 
\colhead{$A_V$ (mag)}} 
\startdata 
none & 2.615 & 0.854 $\pm$ 0.036 & none \\ 
MW & 3.137 & 0.855 $\pm$ 0.058 & 0.00 $\pm$ 0.032 \\ 
LMC & 2.643 & 0.736 $\pm$ 0.129 & 0.067 $\pm$ 0.070 \\ 
SMC & 0.896 & 0.519 $\pm$ 0.104 & 0.139 $\pm$ 0.041 \\ \hline 
\enddata 		              		       
\tablecomments{Spectral energy distribution for UV, optical, and near-IR data only.  We see that the SMC dust is strongly preferred.}   
\end{deluxetable}  
 
 
\begin{deluxetable}{ c  c  c  c  c  c  c } 
\tablewidth{0pc} 
\tablecaption{X-RAY/UV/OPTICAL/NIR SPECTRAL ENERGY DISTRIBUTION\label{tab:sed2}} 
\tabletypesize{\footnotesize} 
 
\tablehead{\colhead{Model} & \colhead{$\Gamma_1$} & \colhead{E$_{\rm bk}$} & \colhead{$\Gamma_2$} & \colhead{$E(B-V)$} & \colhead{$N_{\rm H}$} & \colhead{$\chi^2$/d.o.f.}  
\\ & & keV & & (mag) & ($10^{21}$ cm$^{-2}$) & }
\startdata
PL+MW & 1.99$\pm$0.01 & - & - & 0$^{+8\times 10^{-4}}_{-0}$ & 2.5$^{+1.6}_{-1.3}$ 
& 3.74 \\ 
PL+LMC & 1.99$\pm$0.01 & - & - & 0$^{+8\times 10^{-4}}_{-0}$ & 
2.5$^{+1.6}_{-1.3}$ & 3.74 \\ 
PL+SMC & 1.99$\pm$0.01 & - & - & 0$^{+9\times 10^{-4}}_{-0}$ & 
2.5$^{+1.6}_{-1.3}$ & 3.74 \\ 
BKNPL+MW & 1.6$\pm$0.1 & 0.0019$\pm$0.0002 & 2.03$\pm$0.03 & 
$<$0.02 & 1.4$^{+1.6}_{-{1.3}}$ & 1.72 \\ 
BKNPL+LMC & 1.5$\pm0.1$ & 0.0023$\pm$0.0001 & 2.06$^{+0.03}_{-0.02}$ & 
0.03$\pm$0.01 & 1.8$\pm$1.3 & 1.17 \\ 
BKNPL+SMC & 1.6$\pm$0.1 & 0.0024$^{+0.0002}_{-0.0004}$ & 2.06$^{+0.02}_{-0.03}$ & 
0.02$^{+0.04}_{-0.01}$ & 1.8$^{+1.7}_{-1.3}$ & 1.31 \\ 
BKNPL+MW & $\Gamma_2$-0.5&0.0019$\pm$0.0002 & 2.04$^{+0.03}_{-0.02}$ & 
0.02$\pm$0.01 & 1.3$^{+1.6}_{-{1.2}}$ & 1.73 \\ 
BKNPL+LMC & $\Gamma_2$-0.5 & 0.0023$^{+0.0001}_{-0.0002}$ & 2.06$\pm$0.02 & 
0.028$^{+0.007}_{-0.008}$ & 1.8$^{+1.7}_{-1.3}$ & 1.13 \\ 
BKNPL+SMC & $\Gamma_2$-0.5 & 0.0024$^{+0.0003}_{-0.0005}$ & 2.07$^{+0.05}_{-0.03}$ & 
0.030$^{+0.007}_{-0.014}$ & 1.8$^{+1.7}_{-1.3}$ & 1.30 \\ 
\enddata  
 
\tablecomments{	Results of fits to the near-IR/optical/UV/X-ray SED at 4.26 
days since trigger, with power law and broken power law continuum models  
absorbed by MW-, LMC-, and SMC-like extinction and by X-ray absorption  
assuming solar metallicity. Galactic absorption and extinction were fixed at 
$N_{\rm H,Gal} = 4.8 \times 10^{20}$ cm$^{-2}$ and $E(B-V) = 0.052$ mag,  
respectively, and only intrinsic (at $z = 1.547$) quantities are reported  
in the table. PL = power law (where $\Gamma = 1+ \beta$); BKNPL = broken  
power law; MW = Milky Way extinction curve; LMC = Large Magellanic Cloud  
extinction curve; SMC = Small Magellanic Cloud extinction curve  
\citep[as parametrized by][]{1992ApJ...395..130P}.} 
 
\end{deluxetable}  
 
 
\begin{deluxetable}{ccccc} 
\tablewidth{0pc} 
\tablecaption{ABSORPTION LINES\label{tab:ew}} 
\tabletypesize{\footnotesize} 
\tablehead{\colhead{$\lambda_{obs}$} &\colhead{$W_\lambda^a$} & \colhead{Trans.} & \colhead{$\lambda_{rest}$} & 
\colhead{$z_{abs}$} \\ 
(\AA) & (\AA) & & (\AA) & }  
\startdata 
3316.6&$< 2.766$&    OI 1302  &1302.168&1.5470\\ 
3375.3&$ 3.838\pm0.706$&&&\\ 
3549.9&$< 1.810$&  SiIV 1393 & 1393.755 & \\
3888.5&$< 1.180$&  SiII 1526  &1526.707&1.5470\\ 
3947.1&$ 2.586\pm0.318$&   CIV 1548  &1548.195&1.5495\\ 
&&   CIV 1550  &1550.770&1.5452\\ 
3989.2&$ 1.501\pm0.428$&&&\\ 
4096.7&$< 1.180$& FeII 1608 & 1608.451 & \\
4255.5&$< 1.124$&  AlII 1670  &1670.787&1.5470\\ 

\enddata 
\tablenotetext{a}{Observed equivalent width.} 
\end{deluxetable} 
 
 

\end{document}